\documentclass[fleqn,usenatbib]{mnras}

\usepackage{newtxtext,newtxmath}
\usepackage[T1]{fontenc}
\usepackage{ae,aecompl}

\usepackage{graphicx}	% Including figure files
\usepackage{amsmath}	% Advanced maths commands
\usepackage{amssymb}	% Extra maths symbols

\usepackage[normalem]{ulem}
\usepackage{ragged2e}
\usepackage{xcolor}

\title[]{Inward Bound: The incredible journey of massive black holes as they pair and merge; I. The effect of mass ratio in flattened rotating galactic nuclei.}

\author[F. M. Khan et al.]{Fazeel Mahmood Khan$^{1,2}$, 
	Muhammad Awais Mirza$^{3}$
	and	Kelly Holley-Bockelmann$^{1,4}$
	\\
	$^{1}$Department of Physics and Astronomy, Vanderbilt University, Nashville, TN 37240, USA\\
	$^{2}$Department of Space Science, Institute of Space Technology, Islamabad 44000, Pakistan\\
	$^{3}$Argelander-Institut f\"{u}r Astronomie, Universit\"{a}t Bonn, Auf dem H\"{u}gel 71, Bonn 53121, Germany \\
	$^{4}$Department of Physics, Fisk University, Nashville, TN 37208, USA\\
}
\date{Accepted XXX. Received YYY; in original form ZZZ}

\pubyear{2016}

\begin{document}
	\label{firstpage}
	\pagerange{\pageref{firstpage}--\pageref{lastpage}}
	\maketitle
	
	% Abstract of the paper
	\begin{abstract}

	Understanding how supermassive black holes (SMBHs) pair and merge helps to inform predictions of off-center, dual, and binary AGN, and provides key insights into how SMBHs grow and co-evolve with their galaxy hosts. As the loudest known gravitational wave source, binary SMBH mergers also hold centerstage for the Laser Interferometer Space Antenna (LISA), a joint ESA/NASA gravitational wave observatory set to launch in 2034. Here, we continue our work to characterize SMBH binary formation and evolution through increasingly more realistic high resolution direct $N$-body simulations, focusing on the effect of SMBH mass ratio, orientation, and eccentricity within a rotating and flattened stellar host. During the dynamical friction phase, we found a prolonged orbital decay for retrograde SMBHs and swift pairing timescales for prograde SMBHs compared to their counterparts in non-rotating models, an effect that becomes more pronounced for smaller mass ratios $M_{\rm sec}/M_{\rm prim} = q$. During this pairing phase, the eccentricity dramatically increases for retrograde configurations, but as the binary forms, the orbital plane flips so that it is almost perfectly prograde, which stifles the rapid eccentricity growth. In prograde configurations, SMBH binaries form and remain at comparatively low eccentricities. As in our prior work, we note that the center of mass of a prograde SMBH binary itself settles into an orbit about the center of the galaxy. Since even the initially retrograde binaries flip their orbital plane, we expect few binaries in rotating systems to reside at rest in the dynamic center of the host galaxy, though this effect is smaller as $q$ decreases.

	\end{abstract}

	% Don't make up new ones.

	%%%%%%%%%%%%%%%%%%%%%%%%%%%%%%%%%%%%%%%%%%%%%%%%
	
	%%%%%%%%%%%%%%%%% BODY OF PAPER %%%%%%%%%%%%%%%%%%

	\section{Introduction}\label{sec-intro}
	
	 Over cosmic time, galaxy evolution is punctuated by mergers, especially at high redshift when galaxies assembles~\citep{cand19}. Each merging galaxy injects its supermassive black hole (SMBH) into the remnant, and if the mass of each progenitor galaxy is comparable, the SMBHs will rapidly form a binary at the remnant core  ($q \equiv M_{\rm secondary}/M_{\rm primary} \geq 0.01$) \citep{cal11,Dosopoulou+17,Tremmel+18}. Once in the galaxy core, the subsequent evolution of a SMBH pair is governed by three distinct physical phases: dynamical friction, $3-$body stellar encounters and finally gravitational wave emission \citep{begelman+80}. The stellar environment, morphology and kinematics can profoundly affect the efficiency, onset, and detailed behavior of the SMBHs in each phase, making it non-trivial to map out SMBH merger timescales over the wide variety of galaxy hosts~\citep[see e.g.][]{khan+13, vasiliev+15}.
   
	Let us take a moment to review the physics governing the journey of two separate SMBHs as they merge into one.
	The first phase of SMBH binary evolution, dynamical friction, is outlined in \citet{Chandrasekhar+43}, which describes the trajectory of a heavy point mass moving through a sea of lighter stars. If an infalling SMBH were stripped of its stellar shroud, 
    this description would be appropriate, though in the Chandrasekhar formalism, this stellar sea is infinite, homogeneous, isotropic, and non-rotating. Several recent studies have quantified dynamical friction timescales for SMBHs in the context of a non-rotating galactic potential~\citep{just05,just11,anto12,Dosopoulou+17,Rasskazov+16}, finding timescales of the order of 100 Myr for million solar mass black holes to traverse the final $\sim$ 100 parsecs to binary formation.

    Once the binary forms and hardens, further shrinking the SMBH binary's orbit requires few body interactions that transfer energy and angular momentum to the scattered stars. This few-body scattering phase has been rigorously modeled using a broad range of techniques and underlying assumptions; regardless of the technique, the SMBH binary is either invoked in equilibrium galaxy models \citep{makino+04,berczik+05,berczik+06,mer06,khan+13,vasiliev+15,bor16,gua17,Lez19} or allowed to form self-consistently through galaxy merger simulations \citep{milosavljevic+01,khan+11,preto+11,Gualandris+12,khan+12a,khan+12b,Khan+16,ran17,khan18a,khan18b}. After two decades of work, a converging scenario has emerged: SMBH binaries in the few-body scattering phase continue to shrink at an almost constant rate to a separation where  coalescence via gravitational wave emission is inevitable. SMBH binaries avoid the famous "final parsec problem" by interacting with a large pool of stars on centrophilic orbits in merger remnants having non-spherical geometries. 
    To first order, the semi-major axis evolution in this phase can be approximated by a simple recipe presented in \citet{seskha+15} based on scattering experiments \citep{quinlan+96} and verified by a small set of numerical simulations. 
    
    In the few-body phase, these binaries scour out the centers of their hosts by ejecting a stellar mass that is roughly a few times that of binary\citep{mer06,khan+12a,rantala+18,rantala+19}, hence giving a plausible explanation to the observed flatness of light profiles at centers of giant elliptical galaxies \citep{graham+04,ferrarese+06,kormendy+09,dullo+14}. 
    
    Despite the relatively straightforward behavior of the SMBH binary orbital separation, the eccentricity evolution is much more complex. While eccentricity growth of a {\it hard} binary in a non-rotating stellar environment can be estimated based on scattering experiments \citep{quinlan+96,Sesana+06}, the eccentricity behavior {\it before} the hard binary phase is less well-understood. Indeed, numerical simulations suggest that the eccentricity at binary formation is difficult to predict and the further eccentricity evolution as the binary hardens is even more complex. The apparent stochasticity of eccentricity at binary formation has been reported in several studies \citep{Wang+14,khan18b}. However, recent galaxy merger simulations seem to identify one trend: SMBH binaries formed by galaxy mergers with shallow central mass profiles (with a logarithmic slope, $\gamma \leq$ 1) have a high initial eccentricity ($ e \sim 0.6-0.9$) that grows more eccentric in the hardening phase, while mergers between galaxies with steep cusps ($\gamma \geq 1.5$) produce low eccentricity binaries ($e \sim 0.1-0.3$) which exhibit little subsequent eccentricity evolution \citep{ran17,khan18a}.

    So far, the picture we outlined focused on non-rotating galaxies, but it is clear that rotation affects every stage of SMBH binary evolution. Galaxy rotation is practically ubiquitous; even a classical giant elliptical such as M87, once thought to be a non-rotator, has been shown through exquisite integral field spectroscopic surveys such as ATLAS-3D to rotate to some degree~\citep{Cappellari+07}. Galaxies commonly host rotation-dominated components, such as disks, pseudobulges, and bars, with an appreciable fraction hosting counterrotation\citep[e.g.][]{Rubin70, Sofue+01, Cappellari+16}. Rotation is especially strong for galaxies with stellar masses roughly of order $10^9 -10^{11} M_{\odot}$, and these galaxies tend to host SMBHs in the mass range $10^5 M_{\odot} - 10^7 M_{\odot}$. Because the galaxy mass function rises at lower masses, relatively low mass SMBHs ought to be the most common in the Universe. Mergers between low-mass SMBHs emit gravitational waves that will be detectable with signal-to-noise ratios in the 1000s -- even well into the epoch of reionization -- by the gravitational wave observatory LISA, a joint ESA/NASA mission set to launch in 2034 \citep{amaro+17}. Understanding the evolution of SMBH binaries in galaxy hosts with the appropriate kinematics is therefore an important step to predict LISA observations and to constrain the physics of SMBH evolution from LISA data.
    
    In principle,  SMBH binary evolution
    must be affected by the angular momentum of the stellar background.
    Studies of mostly equal-mass SMBH pairs in rotating galaxy models bear this out \citep{Sesana+11,Gualandris+2012,holley+15,Mirza+17}, showing a few salient differences in evolution compared to their non-rotating counterparts. First, SMBH hardening timescales are approximately 30 percent faster than those in non-rotating systems \citep{holley+15}. In addition, SMBH binaries co-rotating with their stellar host tend to circularise,  whereas those that are counter-rotating become highly eccentric \citep{Sesana+11,holley+15}. Unlike the stable behaviour of the orbital plane in non-rotating models, the binary orbital plane in a rotating galaxy model may flip to become more prograde with respect to the surrounding stellar distribution \citep{Gualandris+2012,Mirza+17}. Surprisingly, the center of mass of the SMBH binary does not settle at the center of a rotating galaxy model, but instead orbits the galaxy center at about the binary influence radius (less than a parsec for a system scaled to the Milky Way, if it were to host a binary \citep{Mirza+17}).

    In this study, we extend our earlier work on the evolution of equal mass SMBH pairs in rotating galaxy models to understand how rotation affects SMBH pairs of different mass ratios. We focus not only the hotly-debated few body scattering phase, but on the dynamical friction phase as well. Our manuscript is organized as follows.  Section \ref{mer-sim} describes our models and initial parameters of our simulations, together with a description of the numerical code and hardware used. In section \ref{BBH-evo}, we show and discuss the results of our simulations. Finally, in section \ref{summary}, we summarize and conclude our study. 
	
	\section{Initial Conditions and Methodology} \label{mer-sim}
	
	 We performed a large set (20) of direct $N-$body simulations, each with one million particles. Our initial galaxy model has an inner logarithmic density profile slope of $\gamma = 1$ \citep{deh93}, and is flattened, with a minor to major axis ratio of 0.8 at the half mass radius. In our model units (MU), the mass of host galaxy is 1 and the central SMBH is 0.005 \citep{har04,Kormendy+13,gra16}. With this mass, the SMBH influence radius (where the enclosed stellar mass is twice that of the SMBH) is 0.05 MU. We used this model in our earlier studies and have shown that equal mass SMBHs merge very efficiently \citep{khan+13,holley+15,Mirza+17}. Rotation was introduced in our galaxy model by flipping the angular momentum of all particles that have an initially negative z-component of angular momentum ($L_z$). 
% 	 \textcolor{red}{
	 This was achieved by swapping the signs of all velocity components
% 	 } 
    \citep{sesana11}. Our runs can broadly be divided into four classes:
    \begin{itemize}
    \item  Non-rotating runs (\textbf{N}-runs) do not have a particular sense of rotation in the host galaxy model.
    
    \item Prograde runs (\textbf{P}-runs), in which the secondary SMBH's initial angular momentum and the galaxy's angular momentum are perfectly aligned in +z direction. 
    
    \item Retrograde runs \textbf{RT}-runs feature a secondary SMBH in a retrograde orbit with an initial inclination tilted $150^{\circ}$ with respect to the host galaxy's $L_z$.
    
    \item In our last set of runs (\textbf{RP}-runs), we set the secondary SMBH on a perfect retrograde orbit *$\theta = 180^{\circ}$ ).
    
    \end{itemize}
    
    %  \textcolor{red}{
    The primary SMBH is initially at rest at the origin, acting as a center for the surrounding stellar cusp
    % }
    . We introduce the secondary SMBHs with a range of masses and velocity orientations to study the effect of mass ratios ($q$) and orbital inclinations on SMBH binary evolution. 
    % \textcolor{red}{
    The secondary SMBH is placed at a separation of 0.5 MU, the scale radius of our model, with $50 \%$ of the circular velocity, resulting in an initial instantaneous eccentricity of approximately 0.6 in each run.
    % }
    We probed five mass ratios ($q = 1, 0.5, 0.25, 0.1, 0.05$) for all four sets of galaxy models described above. SMBHs with mass ratios $q = 1, 0.5 ~ \mathrm{and}~ 0.25$ are expected to form in major galaxy mergers, whereas those with $q \leq 0.1$ are thought to be formed in minor galaxy mergers. Table \ref{tab:runs} describes all our runs and their key parameters.

	\begin{table}
		\caption{Table listing the galaxy models simulated in this study.\label{tab:runs}}
		%\centering
		\begin{tabular}{cccc}
			\hline
			Run & Galaxy and SMBH sense of Rotation  & $\theta$ & Run Time\\
			\hline
			N$_{1.00}$ & No rotation & NA & 0- 80\\ 
			N$_{0.50}$ & No rotation & NA & 0 - 80\\ 
			N$_{0.25}$  & No rotation & NA & 0 - 80\\ 
			N$_{0.10}$ & No rotation & NA & 0 - 100\\
			N$_{0.05}$ & No rotation & NA & 0 - 100\\
			P$_{1.00}$  & Prograde & $0^\circ$ & 0 - 100\\ 
			P$_{0.50}$ & Prograde & $0^\circ$ & 0 - 80\\
			P$_{0.25}$ & Prograde & $0^\circ$ & 0- $\sim$80\\ 
			P$_{0.10}$ & Prograde & $0^\circ$ & 0 - 80\\
			P$_{0.05}$ & Prograde & $0^\circ$ & 0 - 100\\
			RT$_{1.00}$ & Retrograde & $150^\circ$ & 0 - 70\\ 
			RT$_{0.50}$ & Retrograde & $150^\circ$ & 0 - 70\\ 
			RT$_{0.25}$  & Retrograde & $150^\circ$ & 0 - 70\\ 
			RT$_{0.10}$ & Retrograde & $150^\circ$ & 0 - 70\\
			RT$_{0.05}$ & Retrograde & $150^\circ$ & 0 - 100\\
			RP$_{1.00}$ & Retrograde & $180^\circ$ & 0- $\sim$70\\ 
			RP$_{0.50}$ & Retrograde & $180^\circ$ & 0 - 80\\ 
			RP$_{0.25}$  & Retrograde & $180^\circ$ & 0 - 80\\ 
			RP$_{0.10}$ & Retrograde & $180^\circ$ & 0 - 80\\
			RP$_{0.05}$ & Retrograde & $180^\circ$ & 0 - 100\\
%			M & Corotation & $45^\circ$ & 80 - 147\\
			\hline
		\end{tabular}
		
		Column 1: Run configuration. \textbf{N} refers to no-rotation models; \textbf{P} is for prograde SMBHs orbits; \textbf{RT} for SMBHs in retrograde orbits at an inclination of 150$^\circ$ with respect to the galaxy and \textbf{RP} for retrograde SMBH orbits at 180$^\circ$. The subscript denotes the black hole mass ratio.  Column 2: Rotational sense of the binary with respect to the surrounding galaxy.  Column 3: Angle between the SMBH and galaxy angular momenta.  Column 4: Duration of the run in model units.
		
	\end{table}
	
 %   M run is an extension to our A$_{45}$ model--in which SMBHs were %co-rotating at an orientation of 45$^\circ$ w.r.t galaxy rotational %plane--studied in MZA17. The binary was observed to be orbiting in a %wide space about the galactic center. Both of the SMBHs of this %model have now been artificially merged at 70 time units to obtain a %new SMBH in order to explore the final fate of the orbit after the %SMBHs coalescence.
    
    We use a direct summation code, $\phi-GPU$, \citep{Berczik2013} with a 4th order Hermite integrator, to perform our runs on the GPU supported cluster ACCRE\footnote{\url{https://www.vanderbilt.edu/accre/overview/}} at Vanderbilt University. Our softening parameter is set to be $10^{-4}$ for star-star interactions and 0 for SMBH-SMBH interactions. For star-SMBH interactions, its value is $7 \times 10^{-5}$. We stop our runs long after a hard binary (equation \ref{eq:a_h}) is formed and the SMBH binary's semi-major axis and eccentricity have reached a steady phase of evolution.
	
	\section{Results} \label{BBH-evo}
%	Here we discuss findings of our simulations by presenting evolution %of SMBHs in physically distinct phases in sequence of their time %evolution.

    In the following, we concentrate on the behavior of the SMBH binary in separate physical phases.
    
    \subsection{SMBHs pairing -- dynamical friction}
    
        \begin{figure}
        \centering
        \includegraphics[width=1\linewidth]{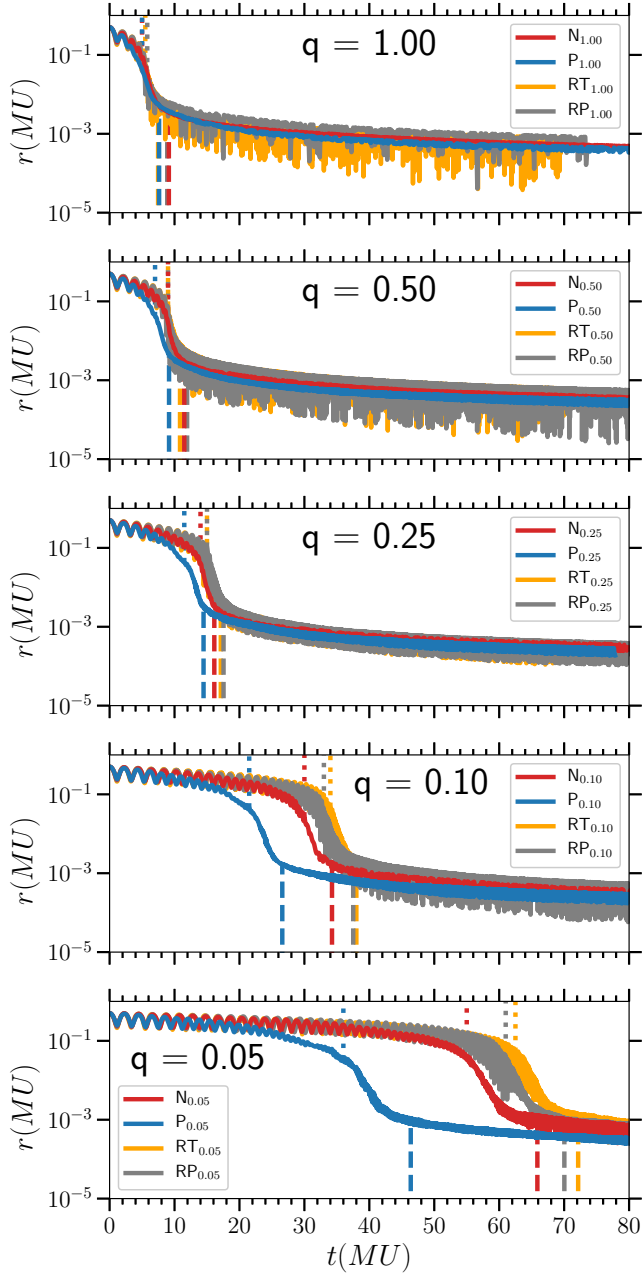}
        \caption{Progression of the separation between the black holes. The dotted vertical lines drawn from the top mark the binary formation time, $t_{bf}$. The dashed vertical lines drawn from bottom indicate the hard binary formation time $t_{h}$, i.e., when a = a$_h$ (equation \ref{eq:a_h}).}
        \label{fig:separation}
    \end{figure}
    
    Figure \ref{fig:separation} shows how the separation shrinks for SMBHs in all runs. Each panel represents a fixed mass ratio for each class of run. The pairing phase is governed by dynamical friction \citep{Chandrasekhar+43} and ends at the binary formation time ($t_{bf}$), which is hard to define in a very exact way. We adopt the convention that a binary is formed once the semi-major axis, as defined by a Keplerian binary (see equation \ref{eq:a}) becomes positive. Here, we neglect the contribution from a bound stellar cusp around the central SMBH, which simplifies the semi-major axis and energy definitions as follows:
    
    \begin{equation}
    a = -\frac{1}{2}\frac{M_{\bullet}}{E_{b}},
    \newline
    \mathrm{where}
    \newline
    E_b = -\frac{M_{\bullet}}{r} + \frac{1}{2}v^2
    \label{eq:a}
    \end{equation}
    is the specific orbital energy, $M_{\bullet}$ is the SMBH binary mass, $r$ is relative separation and $v$ is the relative velocity of SMBHs.
    Binary formation time is shown by the short dashed lines that originate from the top.
    For the equal mass SMBH pair ($q = 1$), the binary formation time $t_{bf}$ is practically unaffected by the kinematics. But as the secondary mass decreases, we witness some clear trends with rotation. Prograde secondaries (\textbf{P}-runs) pair the fastest, while retrograde encounters (\textbf{RT}-runs and \textbf{RP}-runs) take longer. 

    This trend is entirely expected for dynamical friction: in prograde motion, the SMBH and background stars interact longer, resulting in an efficient braking mechanism. This effect is more dramatic for low mass SMBHs where we see a $30 \% $ faster decay time for $q = 0.05$. This sensitivity to rotation on orbit decay has broad implications, especially for low mass SMBHs which have been thought to stall with a standard Chandrasekhar treatment \citep{cal11,Dosopoulou+17,Tremmel+18}, but may pair efficiently in prograde encounters -- or not at all if the SMBH orbit is retrograde.

        \begin{figure}
        \centering
        \includegraphics[width=0.97\linewidth]{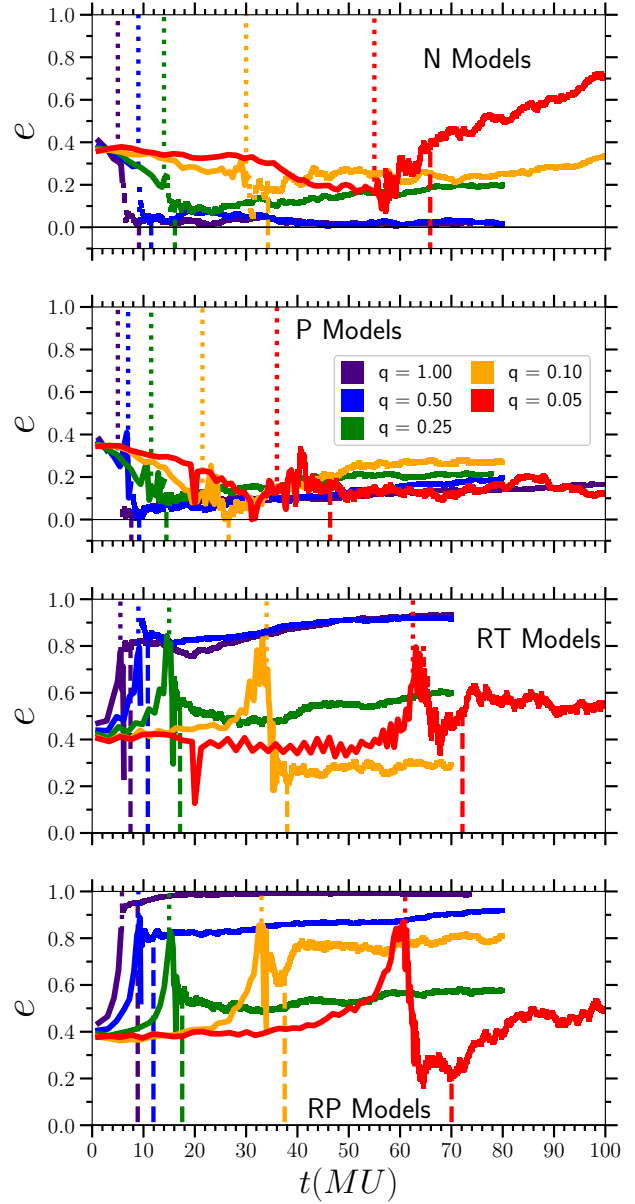}
        \caption{Pattern of eccentricities for the simulation suite; top to bottom: \textbf{N}, \textbf{P}, \textbf{RT} and \textbf{RP} models. Initial SMBH eccentricity has been indicated by a black semi-circle on the vertical axis. The dotted vertical lines (drawn from the top) mark the binary formation time.  On the other hand, the dashed vertical lines from the bottom indicate the hard binary formation time, i.e., when a = a$_h$ which is obtained using eq. \ref{eq:a_h}. Before binary formation, the eccentricity was calculated using $e=\frac{r_a - r_p}{r_a+r_p}$, and afterward we employed equation \ref{eq:eccn-kepler} to obtain smooth realistic plots. The color sequence maps to mass ratio from indigo (I) for equal mass ratio binaries to red (R) for the lowest mass ratio binaries (0.05).\newline
        }
        \label{fig:eccentricity}
    \end{figure}
    
    We found some fascinating eccentricity evolution throughout our suite, which we shall discuss in detail. 
    Figure \ref{fig:eccentricity} shows the complete evolution of eccentricity. First, let us focus only on eccentricity evolution in the pairing phase ( until $t_{bf}$),  when the eccentricity evolution is governed by dynamical friction. We define eccentricity in this phase by peri and apocenter distances (equation \ref{eq:eapoperi}) as the SMBHs are not bound to each other, making the standard definition for a Keplerian binary inapplicable.
        \begin{equation}
    e=\frac{r_a - r_p}{r_a+r_p},
    \label{eq:eapoperi}
    \end{equation}
    where $r_a$ and $r_p$ are apo and pericenter distances of a relative SMBH orbit. 
    
% 	\textcolor{red}{
	
 For \textbf{N} and \textbf{P} runs, we notice that eccentricity decreases gradually in the pairing phase, whereas for the \textbf{RT} and \textbf{RP} runs, $e$ remains constant initially and then rises as we approach binary formation phase.
%  }
	\subsection{Loose Binary -- dynamical friction / $3-$body scattering	}
	
	\begin{figure}
		\centering
		\includegraphics[width=1\linewidth]{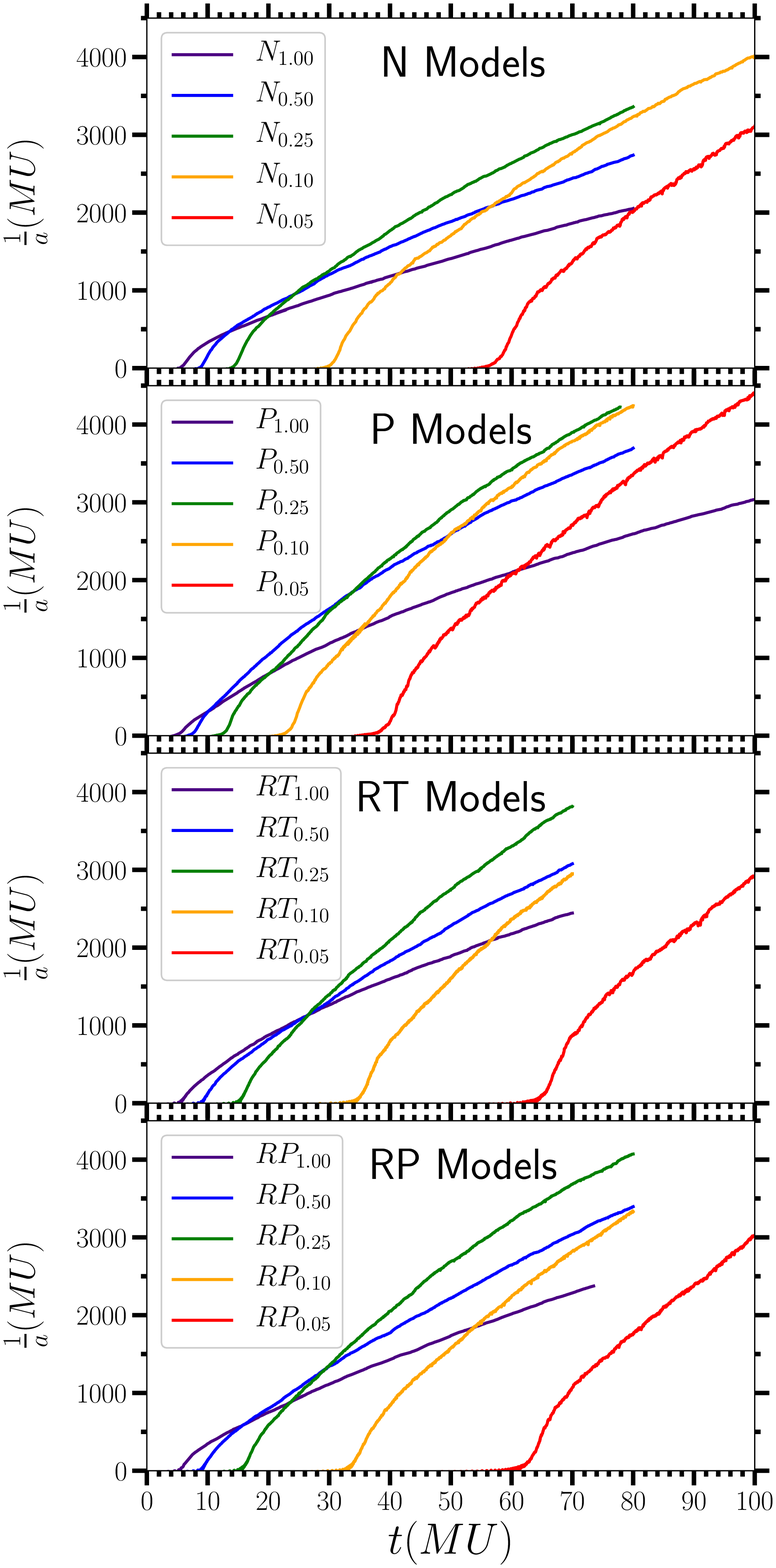}
		\caption{Energy loss of SMBH binaries, as expressed by the inverse semi-major axis.}
		\label{fig:energy-loss}
	\end{figure} 
	
	Once the SMBH sinks to a separation comparable to the influence radius of the two black holes, a loose SMBH binary forms. In addition to dynamical friction, few-body scattering of stellar particles begins to play a role in exchanging energy and angular momentum with the binary's orbit. The combined effect shrinks the separation, r, by almost an order of magnitude in a few time units. As the SMBHs become more tightly bound, the dynamical friction contribution weakens and scattering dominates. For equal mass SMBHs, this phase happens very fast, as both the dynamical friction and few-body scattering scales with SMBH mass. By the same token, SMBH binaries with smaller $q$'s linger in this intermediate phase, which ends at hard binary formation (discussed in detail in the subsequent section and shown by bottom vertical lines in these figures). 
    In the loose binary phase, we see a steep rise in the binary's inverse semi-major axis (Fig. \ref{fig:energy-loss}) which then settles into more linear growth as this phase ends.  
	
		\begin{table*}{}
		\caption{Table of SMBH binary parameters.\label{tab:binary-param}}
		%		\centering
		\begin{tabular}{cccccllcccc}
			\hline
		   Model    & t$_{bf}$ & e$_{bf}$ & t$_{flip}$ & e$_{flip}$ & t$_h$ & e$_h$ & e$_{final}$ &  s   & $\theta^o$ & \.{L}$_{avg}$ (x 10$^{-3}$) \\ \hline
			N$_{1.00}$ &   5.0   &   0.31   &     NA     &     NA     & 9.1  & 0.02  &    0.02     & 23.0 &   0.8   & 35.2                    \\
			N$_{0.50}$ &   9.0   &   0.23   &     NA     &     NA     & 11.5  & 0.03  &    0.02     & 29.0 &   1.9   & 23.2                    \\
			N$_{0.25}$ &   14.0   &   0.22   &     NA     &     NA     & 16.1  & 0.09  &    0.20     & 41.3 &   5.4   & 14.6                    \\
			N$_{0.10}$ &   30.0   &   0.24   &     NA     &     NA     & 34.2  & 0.19  &    0.33     & 55.0 &   5.2   & 7.06                    \\
			N$_{0.05}$ &   55.0   &   0.16   &     NA     &     NA     & 65.9  & 0.44  &    0.71     & 63.5 &   8.1   & 3.74                    \\
			P$_{1.00}$  &   5.0   &   0.16   &     NA     &     NA     & 7.6  & 0.02  &    0.16     & 27.1 &   0.7   & 40.2                    \\
			P$_{0.50}$  &   7.0   &   0.20   &     NA     &     NA     & 9.2  & 0.06  &    0.20     & 40.5 &   4.6   & 28.9                    \\
			P$_{0.25}$  &   11.5   &   0.14   &     NA     &     NA     & 14.5  & 0.09  &    0.22     & 53.8 &   3.5   & 17.7                    \\
			P$_{0.10}$  &   21.5   &   0.10   &     NA     &     NA     & 26.6  & 0.06  &    0.27     & 65.2 &   10.0   & 9.46                    \\
			P$_{0.05}$  &   36.0   &   0.14   &     NA     &     NA     & 46.4  & 0.13  &    0.13     & 76.0 &   12.9   & 5.77                    \\
			RT$_{1.00}$  &   5.5   &   0.76   &    6.0    &    0.63    & 7.5  & 0.82  &    0.93     & 28.4 &   50.3   & 38.2                    \\
			RT$_{0.50}$  &   9.0   &   0.78   &    9.5    &    0.73    & 10.8  & 0.84  &    0.91     & 41.5 &   30.4   & 23.8                    \\
			RT$_{0.25}$  &   15.0   &   0.77   &    14.5    &    0.84    & 17.1  & 0.55  &    0.60     & 57.4 &   20.7   & 14.3                    \\
			RT$_{0.10}$  &   34.0   &   0.67   &    33.5    &    0.66    & 38.0  & 0.27  &    0.29     & 72.4 &   21.1   & 6.34                    \\
			RT$_{0.05}$  &   62.5   &   0.64   &    63.5    &    0.59    & 72.2  & 0.57  &    0.54     & 83.9 &   32.0   & 3.39                    \\
			RT$_{1.00}$ &   5.8   &   0.87   &    54.5    &    0.99    & 8.9  & 0.95  &    0.98     & 28.8 &   81.9   & 37.8                    \\
			RP$_{0.50}$ &   9.0   &   0.83   &    9.5    &    0.81    & 11.9  & 0.81  &    0.92     & 41.9 &   04.9   & 24.7                    \\
			RP$_{0.25}$ &   15.0   &   0.78   &    15.5    &    0.80    & 17.5  & 0.53  &    0.57     & 53.8 &   15.5   & 14.2                    \\
			RP$_{0.10}$ &   33.0   &   0.75   &    33.0    &    0.75    & 37.5  & 0.76  &    0.81     & 65.0 &   15.2   & 6.67                    \\
			RP$_{0.05}$ &   61.0   &   0.76   &    60.5    &    0.83    & 70.0  & 0.36  &    0.49     & 68.3 &   12.9   & 3.62                    \\ \hline
			
		\end{tabular}
		
		\justify
		
		\textit{Notes.} Column 1 (Model): Model name. Column 2 (t$_{bf}$): Time of binary formation. Column 3 (e$_{bf}$): Eccentricity at binary formation time. Column 4 (t$_{flip}$): Time of binary plane flip, i.e., at $\theta$ = 90. Column 5 (e$_{flip}$): Eccentricity at binary plane flip. Column 6(t$_h$): Time of hard binary formation. Column 7(e$_h$): Eccentricity at hard binary formation. Column 8 (e$_{final}$): Final eccentricity of the binaries (at the time when simulations stop, mentioned in Table \ref{tab:runs}). Column 9 (s): Hardening rate of the binaries from 40 to 70 time units, except \textbf{N}$_{0.0{5}}$, \textbf{RT}$_{0.05}$ and \textbf{RP}$_{0.05}$, for which the time interval is from 70 to 80 time units due to the late formation of the binary. Column 10 ($\theta$): Angle (in degrees) between the angular momenta of the binary and that of the galaxy at the time simulations are stopped. Column 11 (\.{L}$_{avg}$): Average total angular momentum loss until the time the binary forms given as multiples of 10$^{-3}$. 
		
	\end{table*}

    Eccentricity behaviour in this intermediate phase is quite diverse and merits a thorough discussion. Once the SMBHs form a binary, we define eccentricity by a Keplerian orbit,
    \begin{equation}
        e=\sqrt{1+\frac{2E_b h^2}{\mu^2}},
        \label{eq:eccn-kepler}
    \end{equation}
    (\noindent where h is the specific relative angular momentum and $\mu$ = G(m$_1$ + m$_2$)).
    \begin{figure}
        \centering
        \includegraphics[width=1\linewidth]{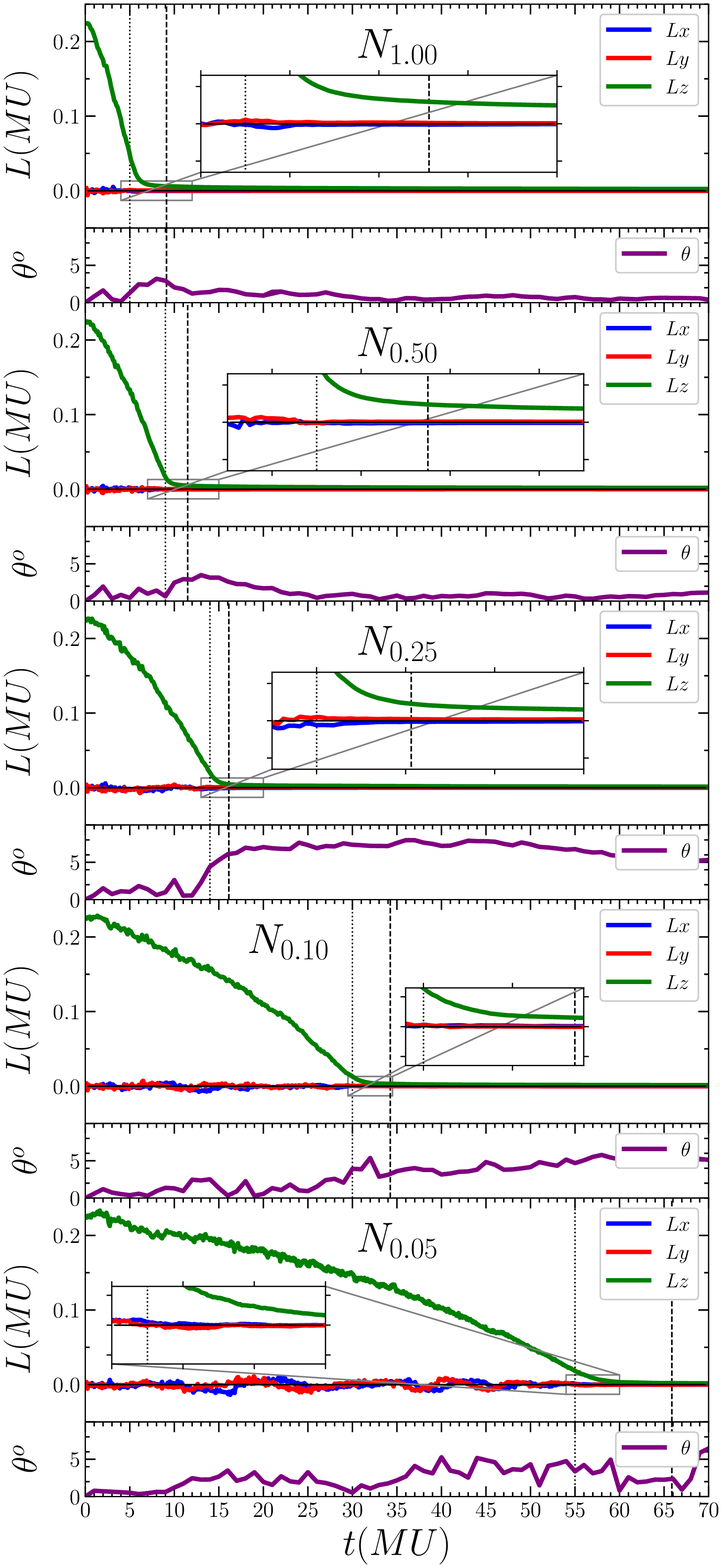}
        \caption{Evolution of angular momentum components (Lx, Ly, Lz) and the angle between the binary and galaxy angular momenta ($\theta$) for \textbf{N} runs. Top to bottom: \textbf{N}$_{1.00}$, \textbf{N}$_{0.50}$, \textbf{N}$_{0.25}$, \textbf{N}$_{0.01}$ and \textbf{N}$_{0.05}$. The dotted vertical line marks the binary formation time, while the dashed vertical line indicates the hard binary formation time. An inset yields a better view at the time when the majority of angular momentum has depleted.  }
        \label{fig:angularRF}
    \end{figure}
     
     In the \textbf{N} models, eccentricity remains almost constant and nearly circular, except for $q = 0.05$. For \textbf{N}${_{0.05}}$, eccentricity begins to rise soon after binary formation and continues growing through hard binary formation and into the hard binary phase. The mechanism of pumping up eccentricity for small $q$'s is consistent with \citet{Iwasawa+11} and is caused by the preferred ejection of stars on co-rotating orbits by an SMBH binary.
     This preferred ejection process happens for all secondary masses, but low mass secondaries are particularly vulnerable to the resulting perturbation by the dominating counter-rotating stellar stream.  
     The secondary's perturbation causes counter-rotating stars to flip angular momentum to co-rotating before being ejected by binary; this requires an angular momentum loss from binary's orbit and hence an increase in eccentricity. Our results show that this mechanism of eccentricity growth is effective only for binaries in non-rotating models with $q \leq 0.02$ and not for SMBH binaries with $q \geq 0.1$.
     The initial angular momentum of the secondary's orbit is in $L_z$ for \textbf{N} runs, and it stays that way during the full duration of our simulations (Fig. \ref{fig:angularRF}). Notice that the orbit has a very stable orientation, staying within $5^{\circ}$ for all runs.   
     
         \begin{figure}
        \centering
        \includegraphics[width=1\linewidth]{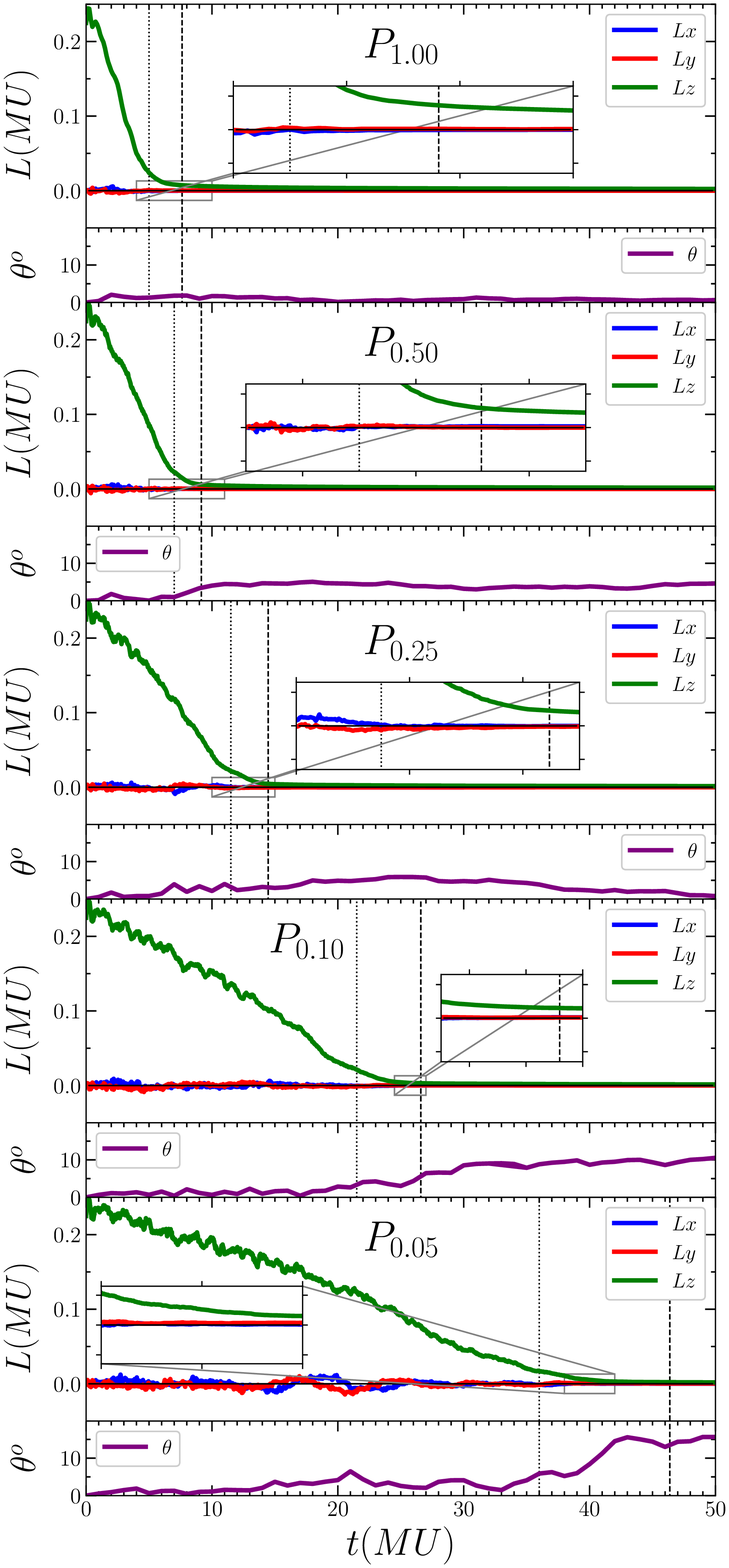}
        \caption{Evolution of the angular momentum components (Lx, Ly, Lz) and the angle between the binary and galaxy angular momenta ($\theta$) for \textbf{P} runs. Top to bottom: \textbf{P}$_{1.00}$, \textbf{P}$_{0.50}$, \textbf{P}$_{0.25}$, \textbf{P}$_{0.01}$ and \textbf{P}$_{0.05}$. The vertical lines and inset have been explained in the caption of Fig. \ref{fig:angularRF}.}
        \label{fig:angularO}
    \end{figure}

    We see low eccentricity in all \textbf{P} models. Here again, the SMBHs orbital plane (Fig. \ref{fig:angularO}) is very stable, strongly aligned with galaxy's rotation plane. 
     
\begin{figure}
        \centering
        \includegraphics[width=1\linewidth]{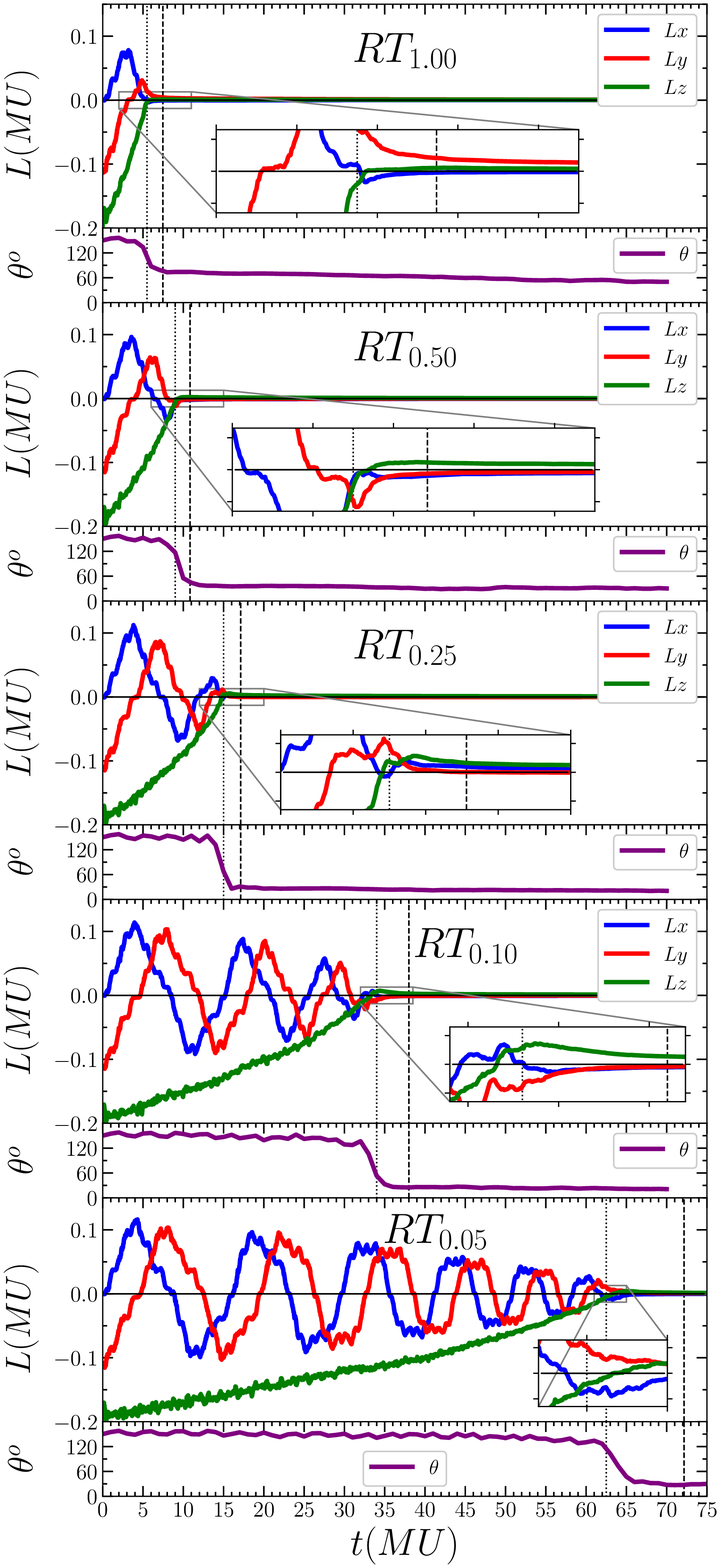}
        \caption{Evolution of the angular momentum components (Lx, Ly, Lz) and angle between binary and galaxy angular momenta ($\theta$)  for N runs. Top to bottom: N$_{1.00}$, N$_{0.50}$, N$_{0.25}$, N$_{0.01}$ and N$_{0.05}$.  The vertical lines and inset have been explained in the caption of Fig. \ref{fig:angularRF}.}
        \label{fig:angularN}
    \end{figure}

\begin{figure}
        \centering
        \includegraphics[width=1\linewidth]{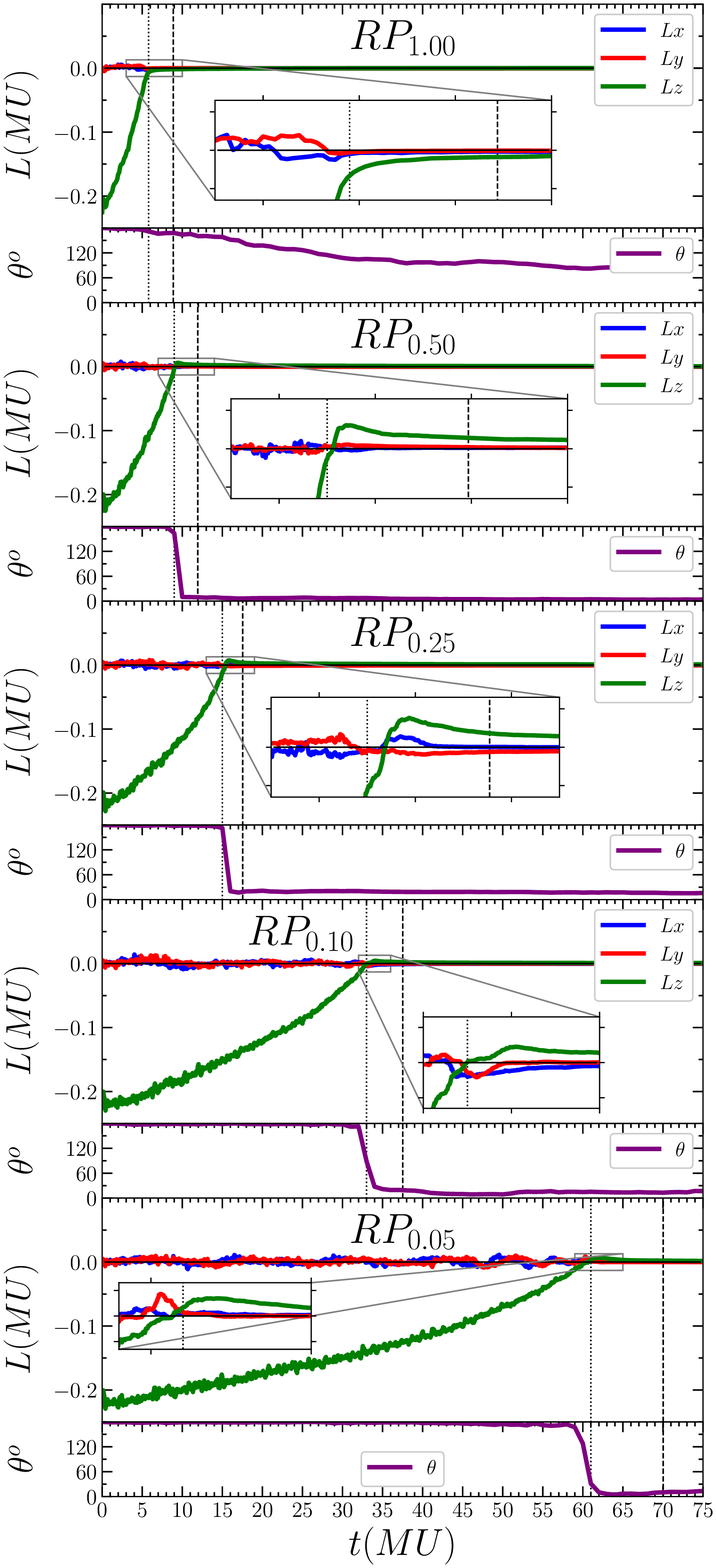}
        \caption{Evolution of the angular momentum components (Lx, Ly, Lz) and angle between binary and galaxy angular momenta ($\theta$) for \textbf{RP} runs. Top to bottom: \textbf{RP}$_{1.00}$, \textbf{RP}$_{0.50}$, \textbf{RP}$_{0.25}$, \textbf{RP}$_{0.01}$ and \textbf{RP}$_{0.05}$. The vertical lines and inset have been explained in the caption of Fig. \ref{fig:angularRF}.}
        \label{fig:angularNP}
    \end{figure}

     For our models hosting SMBHs on retrograde orbits (\textbf{RT} and \textbf{RP}), eccentricities are generally high ($e \simeq 0.5 - 0.99$). Almost all binaries form with high eccentricity ($e \sim 0.8$) and then some reach even higher values, while others begin to circularise. Here the trend is that for more unequal mass SMBHs, the eccentricity drops immediately after it reaches a peak value, but for more equal masses, eccentricity continues to grow ($e \sim 0.9-0.99$). This is consistent with our earlier studies \citep{holley+15,Mirza+17}, where we find high eccentricities ($e \sim 0.9 - 1.0$) for counter-rotating equal mass binaries. Here, we extended these findings to smaller mass ratios. To look more into the eccentricity behaviour in different runs of these model classes, we analyse the angular momentum evolution (Figs. \ref{fig:angularN}, \ref{fig:angularNP}) of SMBHs for each run in this phase. 
     
\begin{figure}
        \centering
        \includegraphics[width = \linewidth]{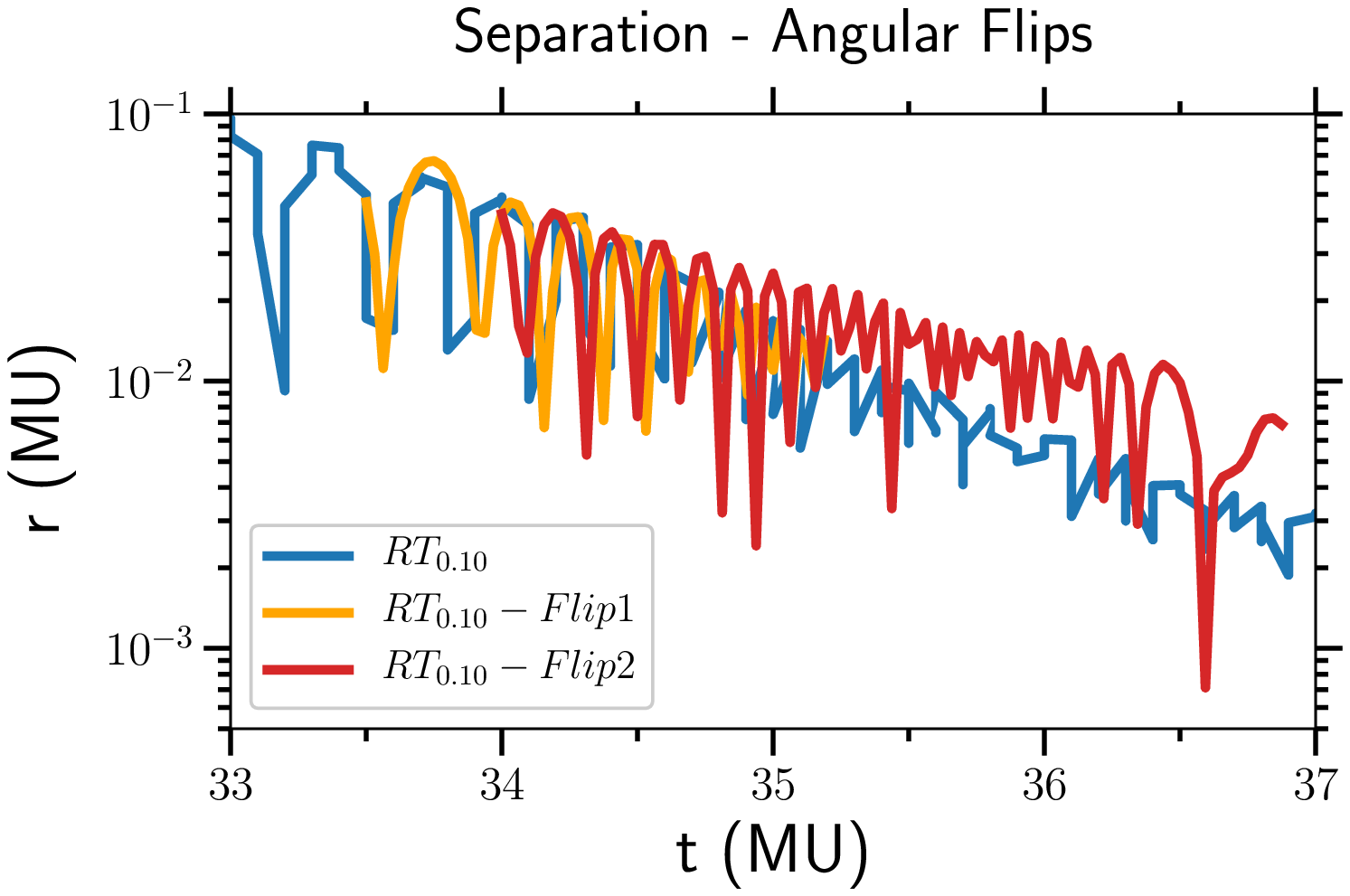}
        \includegraphics[width = \linewidth]{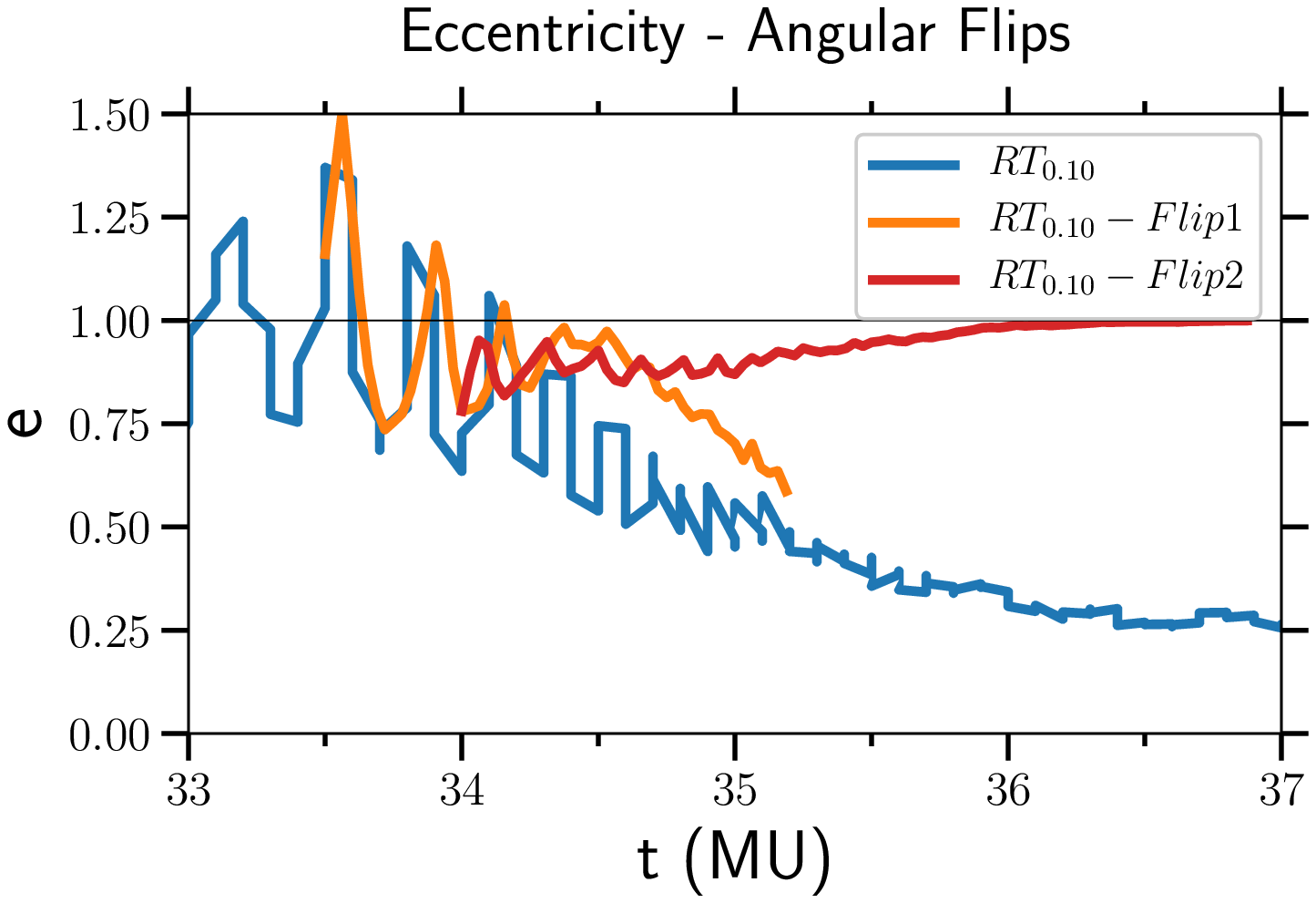}
        \includegraphics[width = \linewidth]{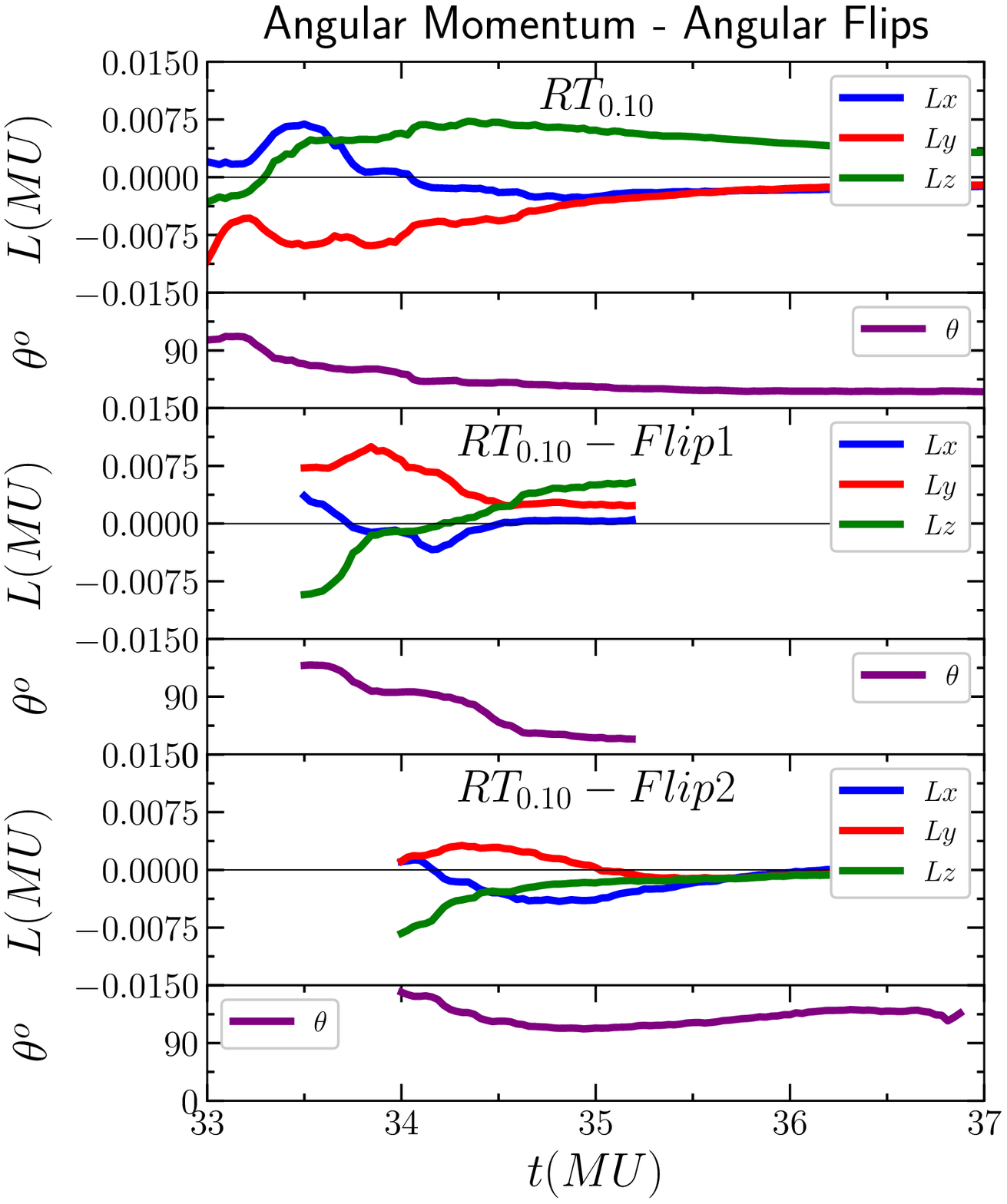}
        \caption{From top to bottom: Relative separation $R$, eccentricity $e$, angular momentum $L$ and angular $\theta$ evolution of run \textbf{RT}$_{0.1}$.  }
        \label{fig:flip_sep}
    \end{figure}

    In analyzing the angular momentum (Figs. \ref{fig:angularRF} - \ref{fig:flip_sep}), we demonstrate that the leap of eccentricity we witness near binary formation is due to the orbital plane abruptly reorienting itself from retrograde to prograde (see Figs \ref{fig:angularN} and \ref{fig:angularNP} in particular). 

    This typical reorientation of counter-rotating binaries to co-rotating states is thought to be because of the strong torques from ejecting stars in the neighborhood \citep{Sesana+11}. After the ascent, the eccentricity plummets for lower mass-ratio SMBHs during the loose binary hardening phase, returning to moderate eccentricity values. We find that it is the secondary SMBH that flips its angular momentum at apo-center (which is $\sim$ of influence radius at binary flip time $t_{flip}$, see Tab. \ref{tab:binary-param}). Soon after flipping the orbital plane, the binary tends to circularise, as is the case for prograde orbits. Runs with $q = 1, 0.5$ are an exception to this; perhaps its difficult for stellar interactions to circularise a more massive binary, as the stellar reservoir is constantly diminishing due to core scouring by the binary. This circularization phase ends at around the hard binary formation time.
    
    To see how the behaviour of eccentricity would be if the orbit remained retrograde, we artificially reverse the SMBH binary angular momentum (returning it to retrograde) for one of these runs - \textbf{RT}$_{0.1}$ at time $t_{flip}$ (Fig. \ref{fig:flip_sep}). The eccentricity remains higher for some time, but again falls as the binary plane flips to co-rotation again. We flip the orbital plane once more, and it remains counter-rotating with a subsequent rise in eccentricity through the end of the run.  We conclude that binary's orbital plane flips at about binary formation time, but if the SMBH binary can remain counter-rotating, it will reach an extremely high eccentricity for all mass ratios. This effect needs further investigation to verify the effect and to understand how it depends on the degree of rotation of surrounding stellar distribution.

	\subsection{Hard binary - few-body scattering}	
	
	Once the binary semi-major axis shrinks to $a_h$, $3-$body scattering preferentially extracts energy, shrinking the orbit further -- at this separation, a hard binary has formed.  
    
    \begin{equation}
    a_h = \frac{r_h}{4} \frac{q}{(1 + q)^2}
    \label{eq:a_h}
    \end{equation}
    Where r$_h$ is the influence radius and q is the black hole mass-ratio.
    
    The eccentricity in the hard binary phase grows gradually, as expected \citep{sesana+10,khan+12b}. 
    The SMBH binary separation (Fig. \ref{fig:separation}) and semi-major axis (Fig. \ref{fig:energy-loss}) shrink at an almost constant rate. We fit a straight line to the inverse semi-major axis to determine the slope, which is the  hardening rate $s$ (Table \ref{tab:binary-param}). Hardening rates are systematically higher for binaries with smaller $q$.
    Furthermore, rotating (\textbf{P}, \textbf{RT} and \textbf{RP}) models are more effective in extracting energy from the binary orbit than non-rotating ones at a given $q$. As a reminder, despite the secondary starting on a counter-rotating orbit in models \textbf{RT} and \textbf{RP}, they end up in co-rotation before the binary reaches the hard-binary phase where we calculate $s$. So essentially, all three cases of rotation end up with hosting co-rotating binaries, hence we do not see difference in hardening rates.
    This has broad implications for strongly rotating bulges and nuclear star clusters: we expect prograde binaries with boosted hardening rates.  
    
    Because the earlier loose binary phase boosted the eccentricity in counter-rotating models, especially for equal masses, the increase in hardening rate found in this stage compounds, resulting in a much more rapid transition into the gravitational wave regime. Its worth remembering, however, that the dynamical friction significantly delay for a retrograde SMBH if its mass ratio is too small $q < 0.1$. Note that these conclusions are true only for fully rotating models. For partial rotation, the scenarios can change and need further investigation.

	\subsection{Binary Center of Mass Trajectory}	
		
	\begin{figure}
		\centering
		\includegraphics[width=1\linewidth]{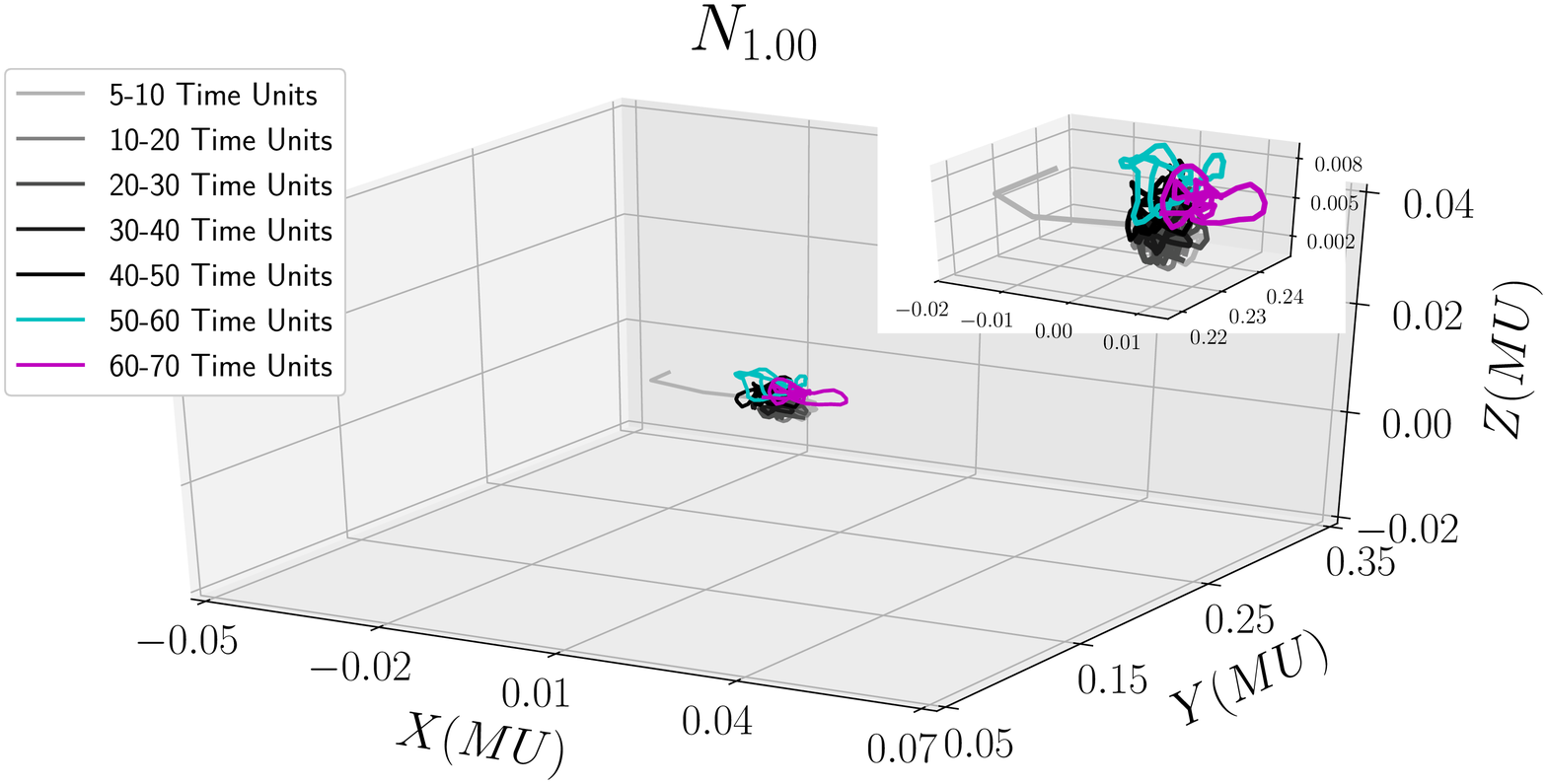}
		\includegraphics[width=1\linewidth]{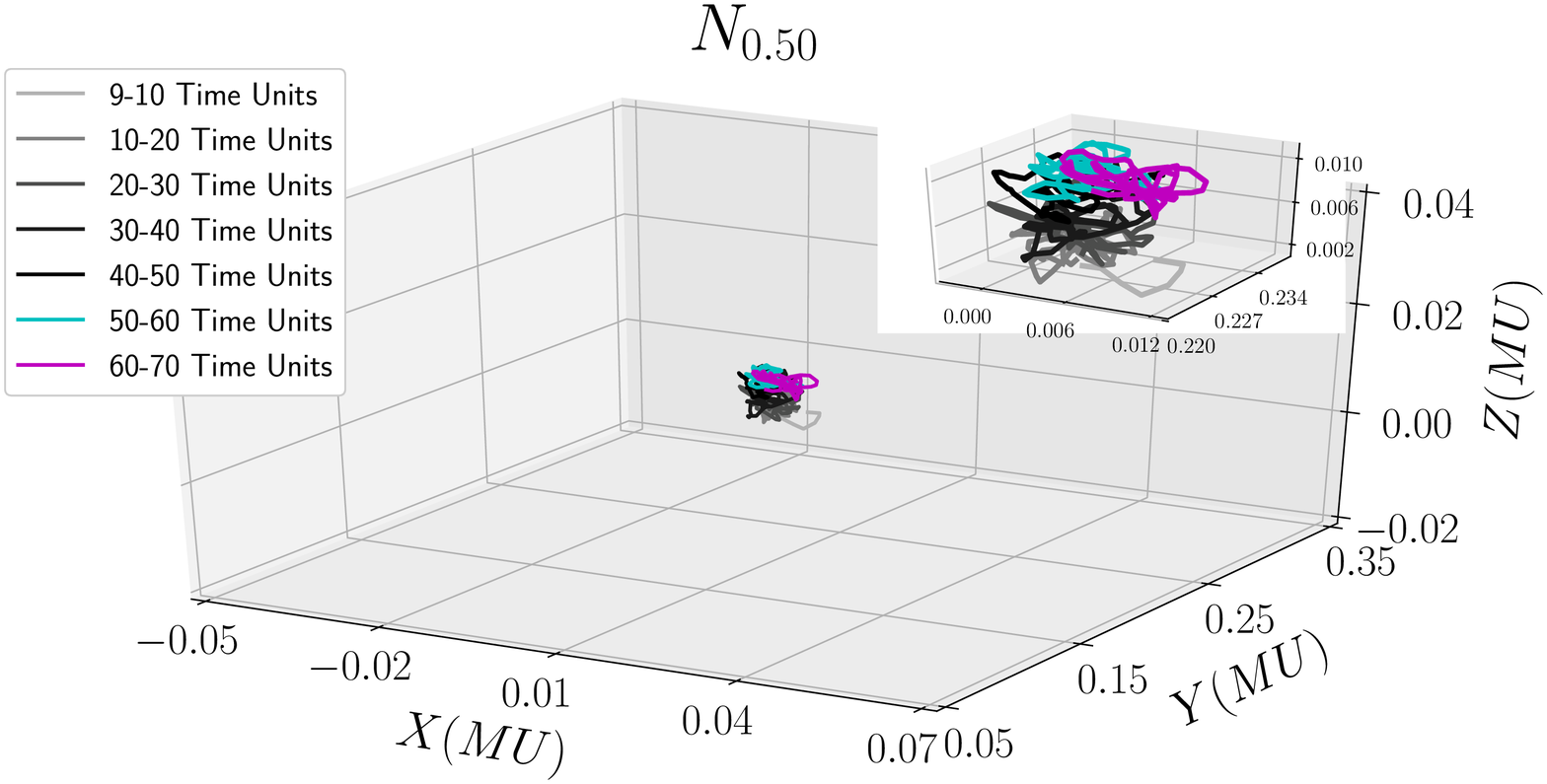}
		\includegraphics[width=1\linewidth]{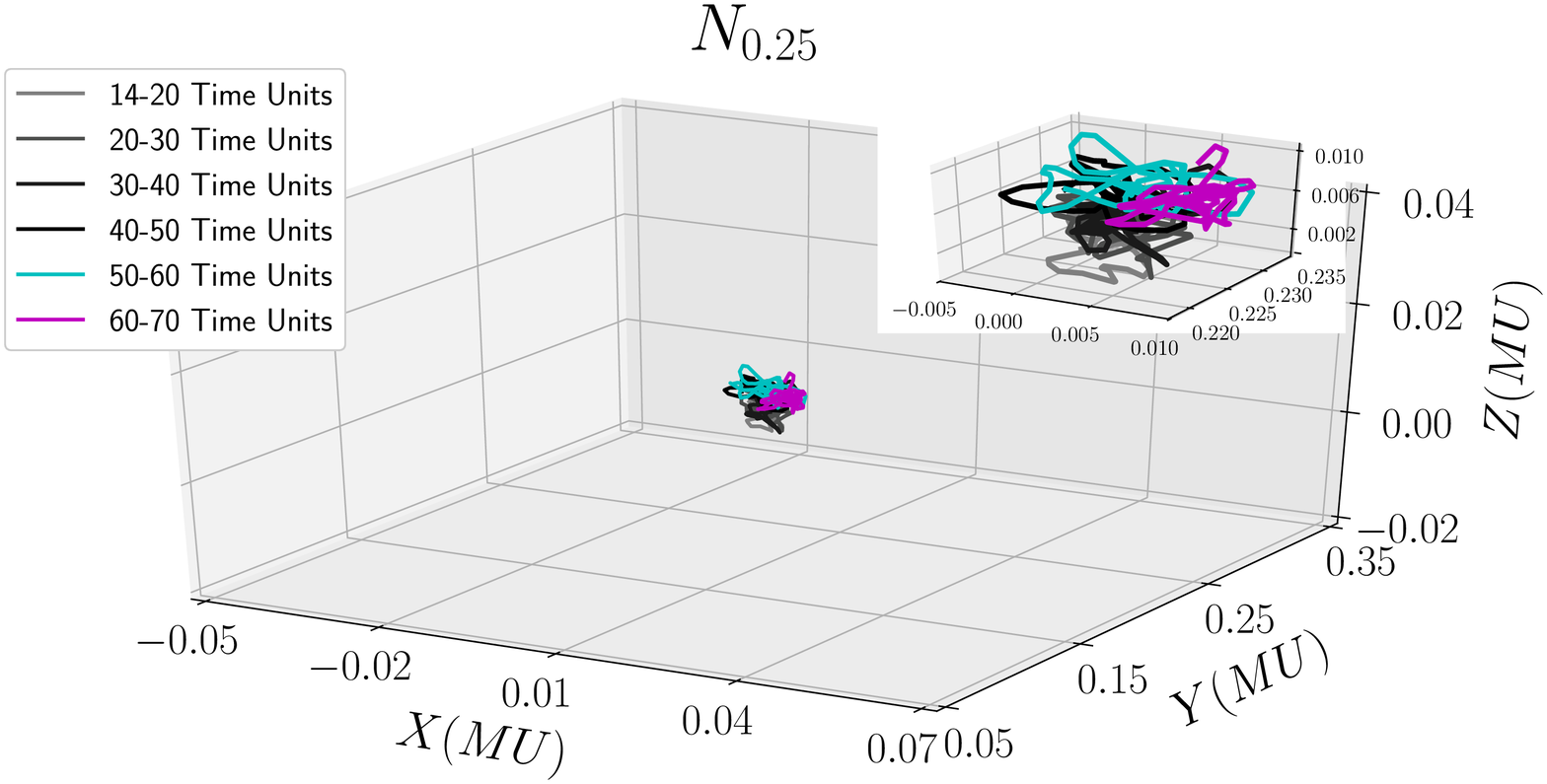}	
		\includegraphics[width=1\linewidth]{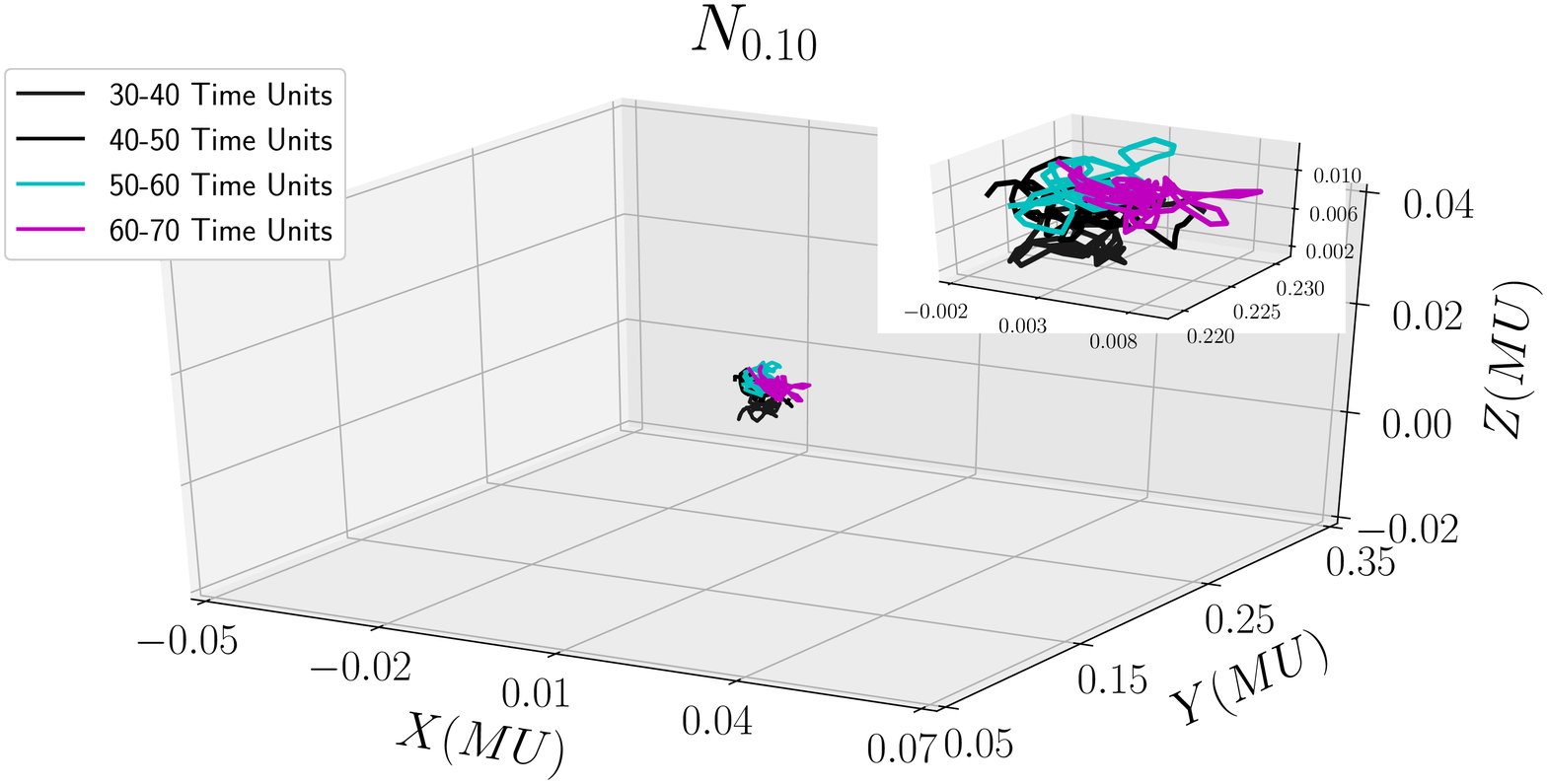}			
		\includegraphics[width=1\linewidth]{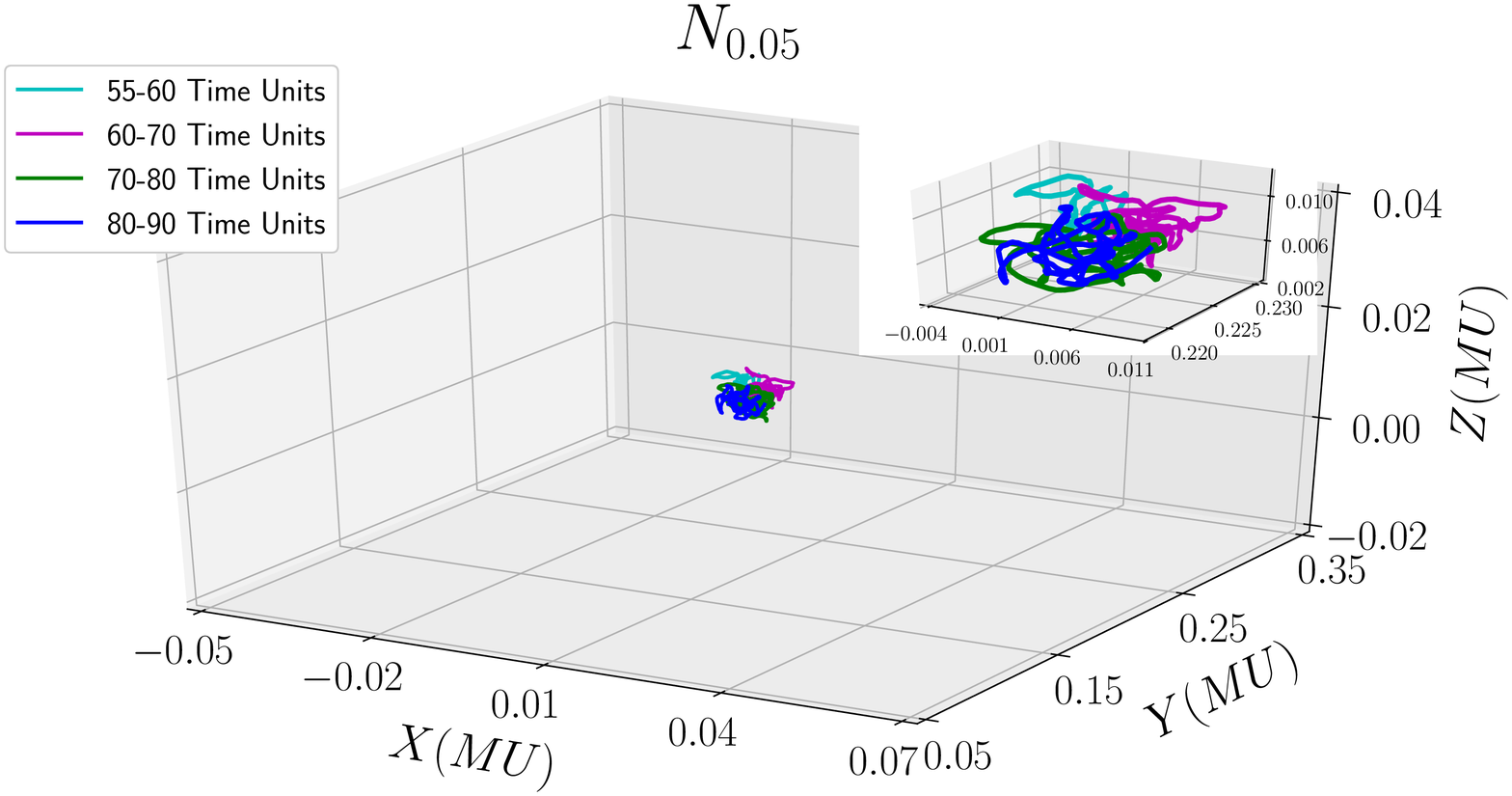}	 
		\caption{Motion of the binary SMBH center of mass for N$_{1.00}$ (top) to N$_{0.05}$ (bottom). To infer relative size of the orbit, the figure scale has been fixed for all \textbf{N}, \textbf{P}, \textbf{RT} and \textbf{RP} suites.}		
		\label{fig:comRF}
	\end{figure}
	
	\begin{figure}
		\centering
		\includegraphics[width=1\linewidth]{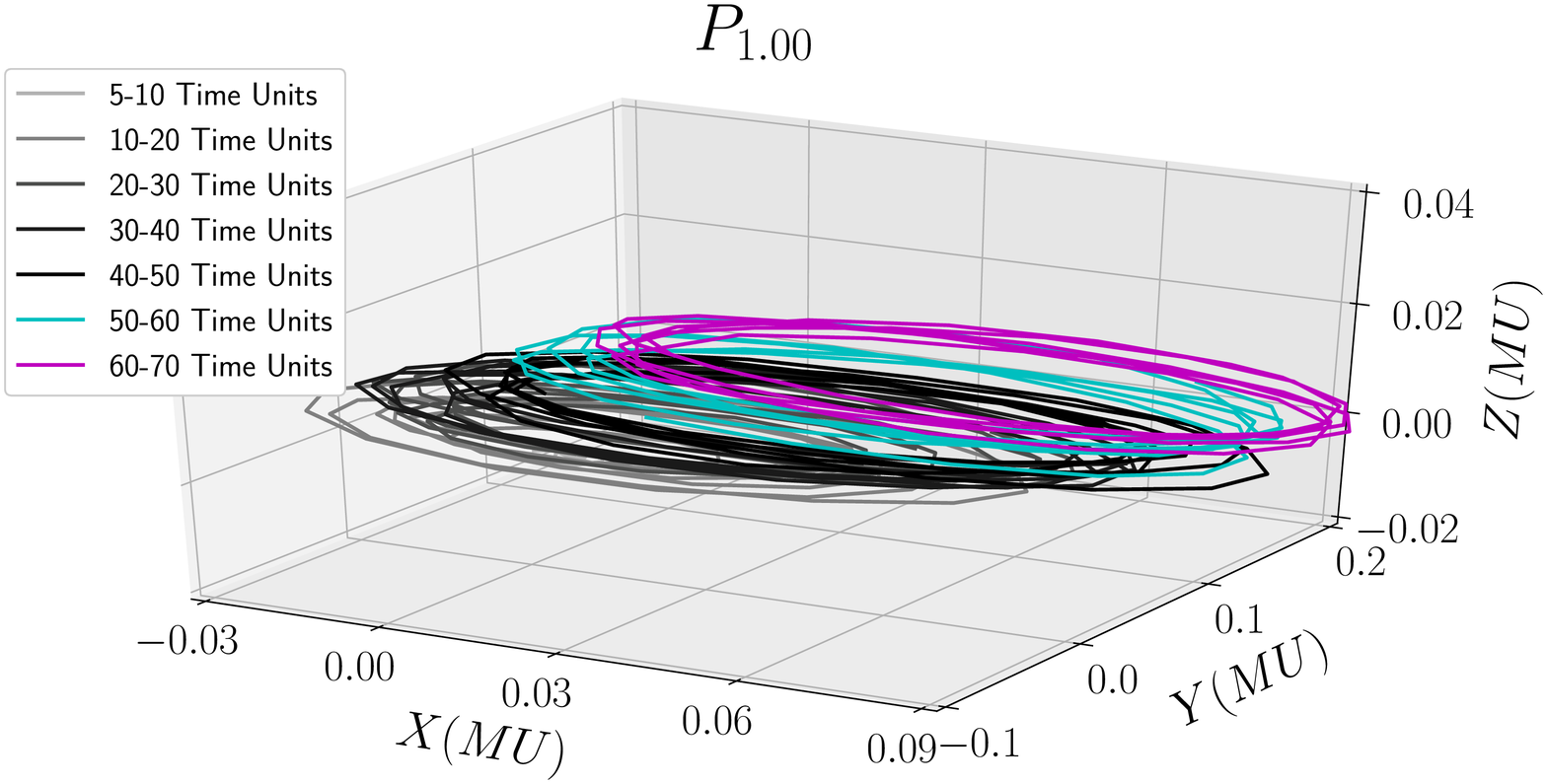}
		\includegraphics[width=1\linewidth]{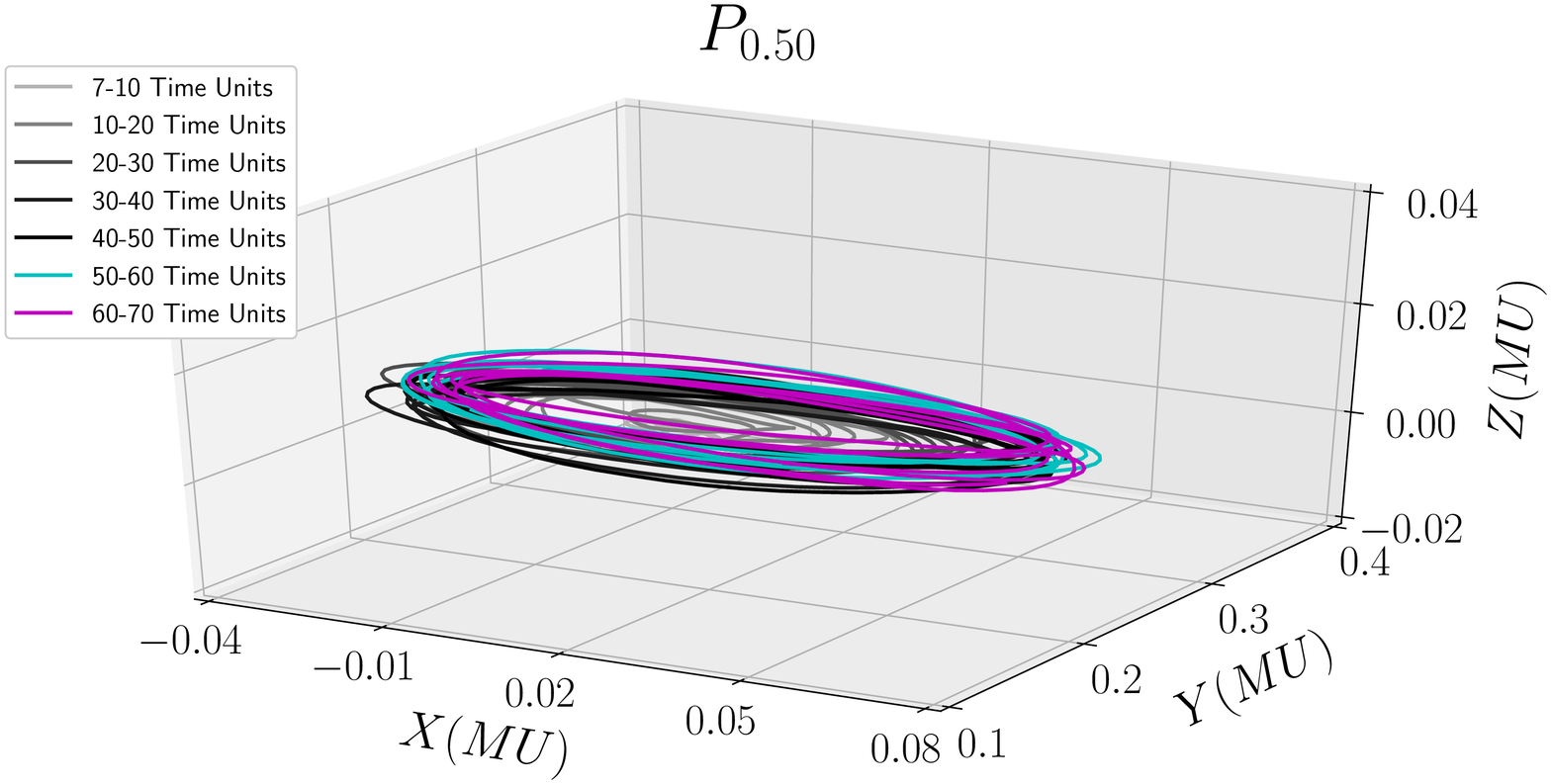}
		\includegraphics[width=1\linewidth]{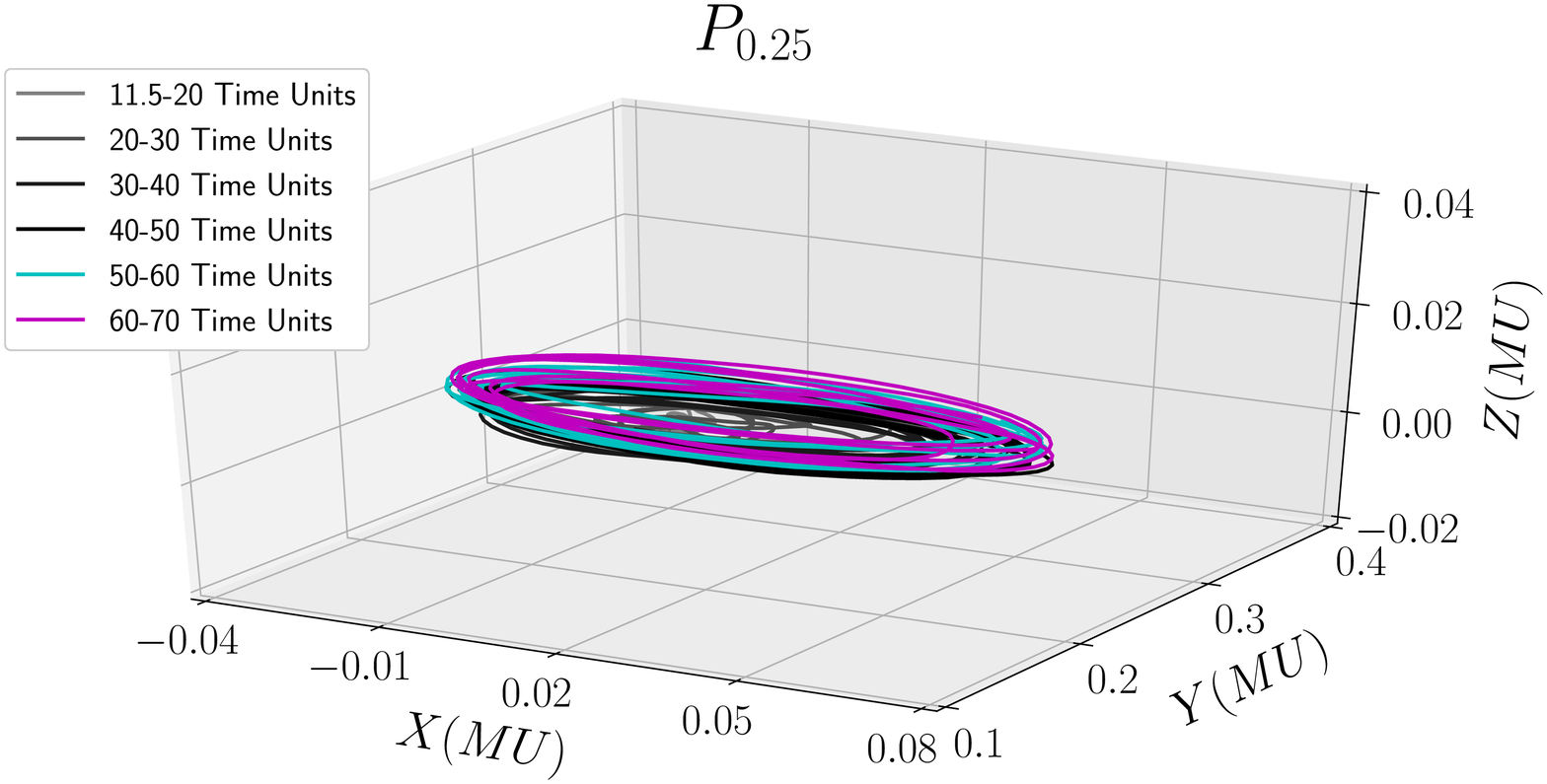}	
		\includegraphics[width=1\linewidth]{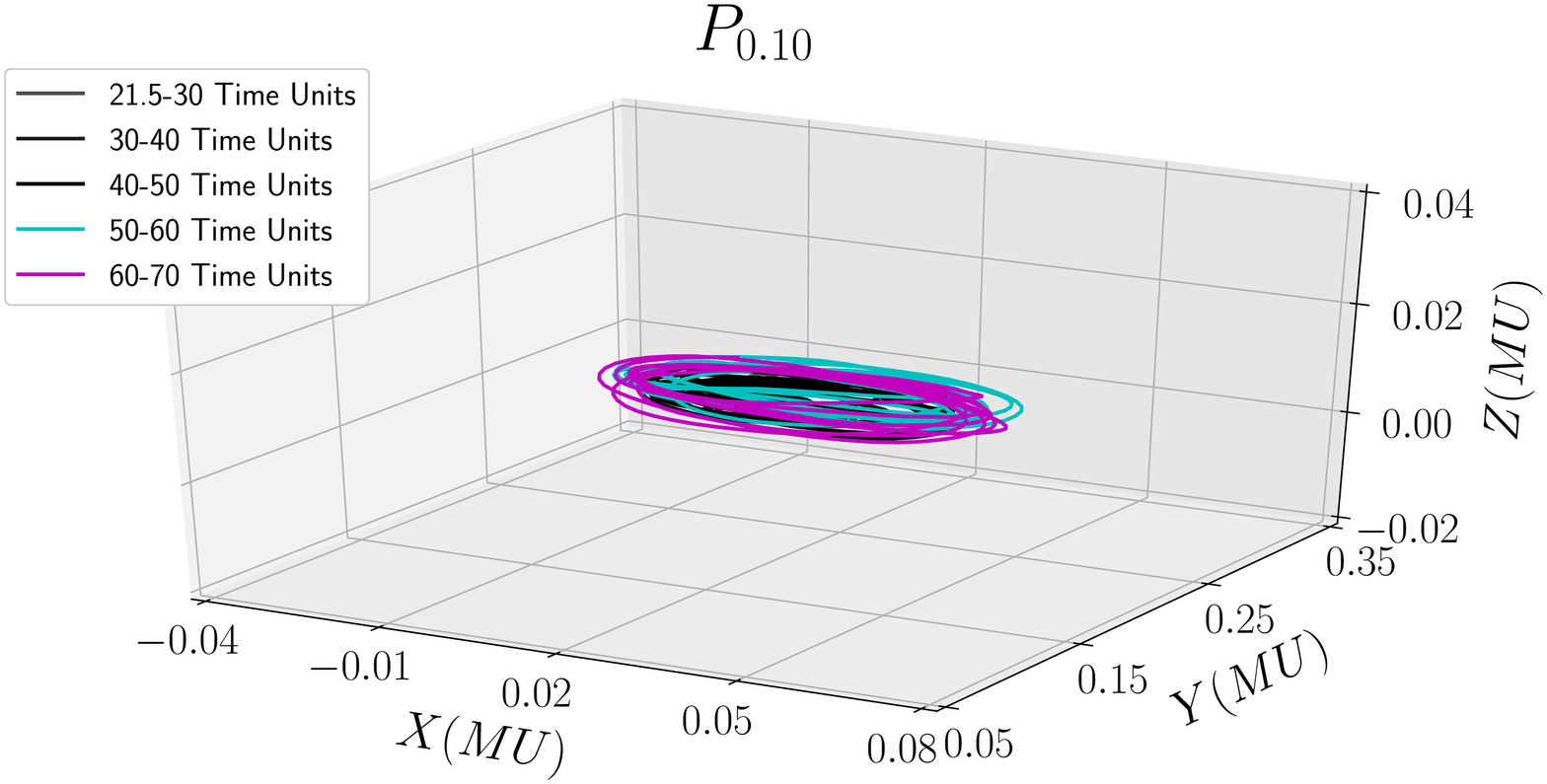}
		\includegraphics[width=1\linewidth]{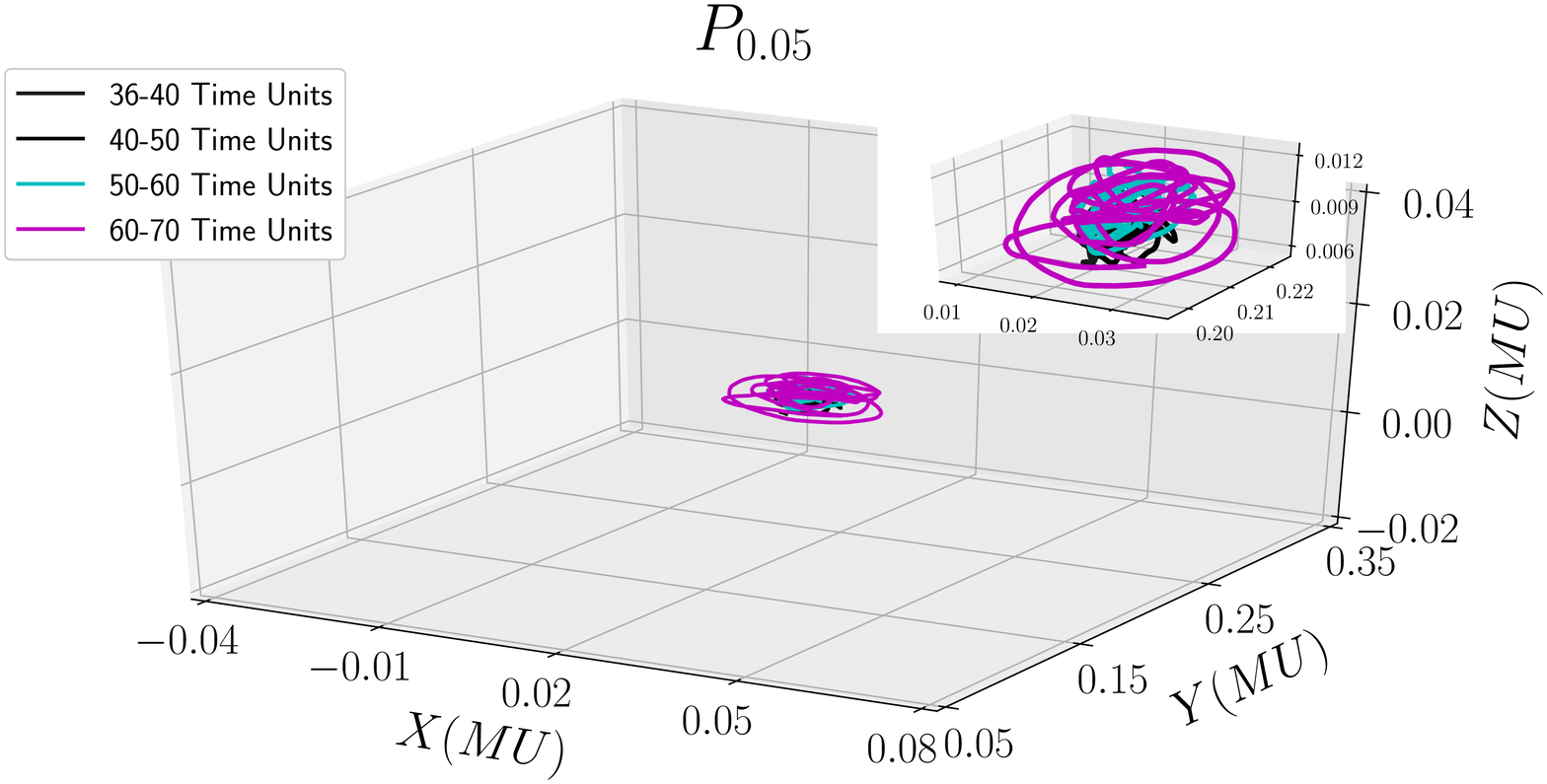}
		\caption{Motion of the binary SMBH center of mass for P$_{1.00}$ (top) to P$_{0.05}$ (bottom). The axes are fixed as discussed in the caption of Fig. \ref{fig:comRF}.}		
		\label{fig:comO}
	\end{figure}
	
	\begin{figure}
		\centering
		\includegraphics[width=1\linewidth]{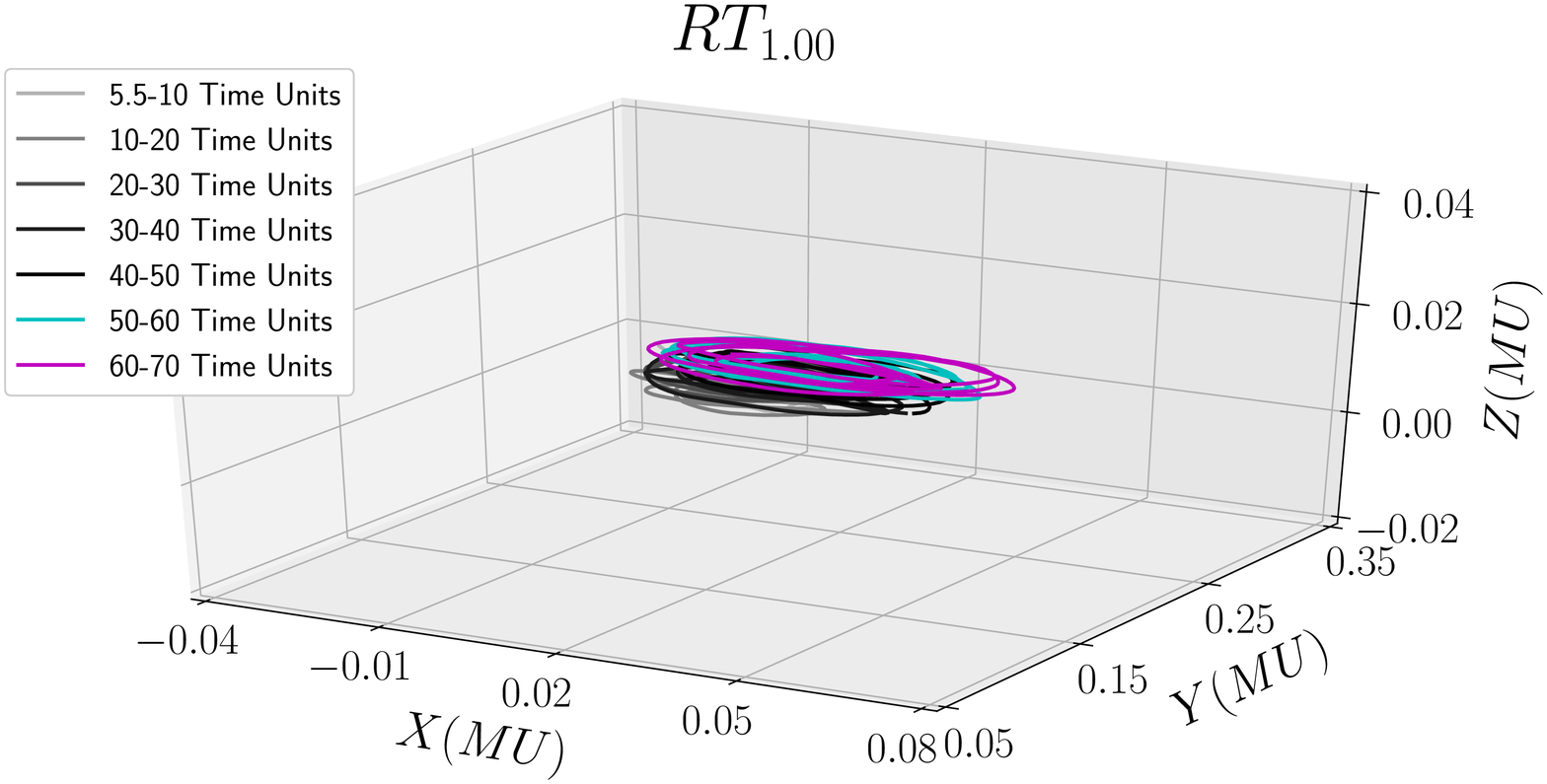}
		\includegraphics[width=1\linewidth]{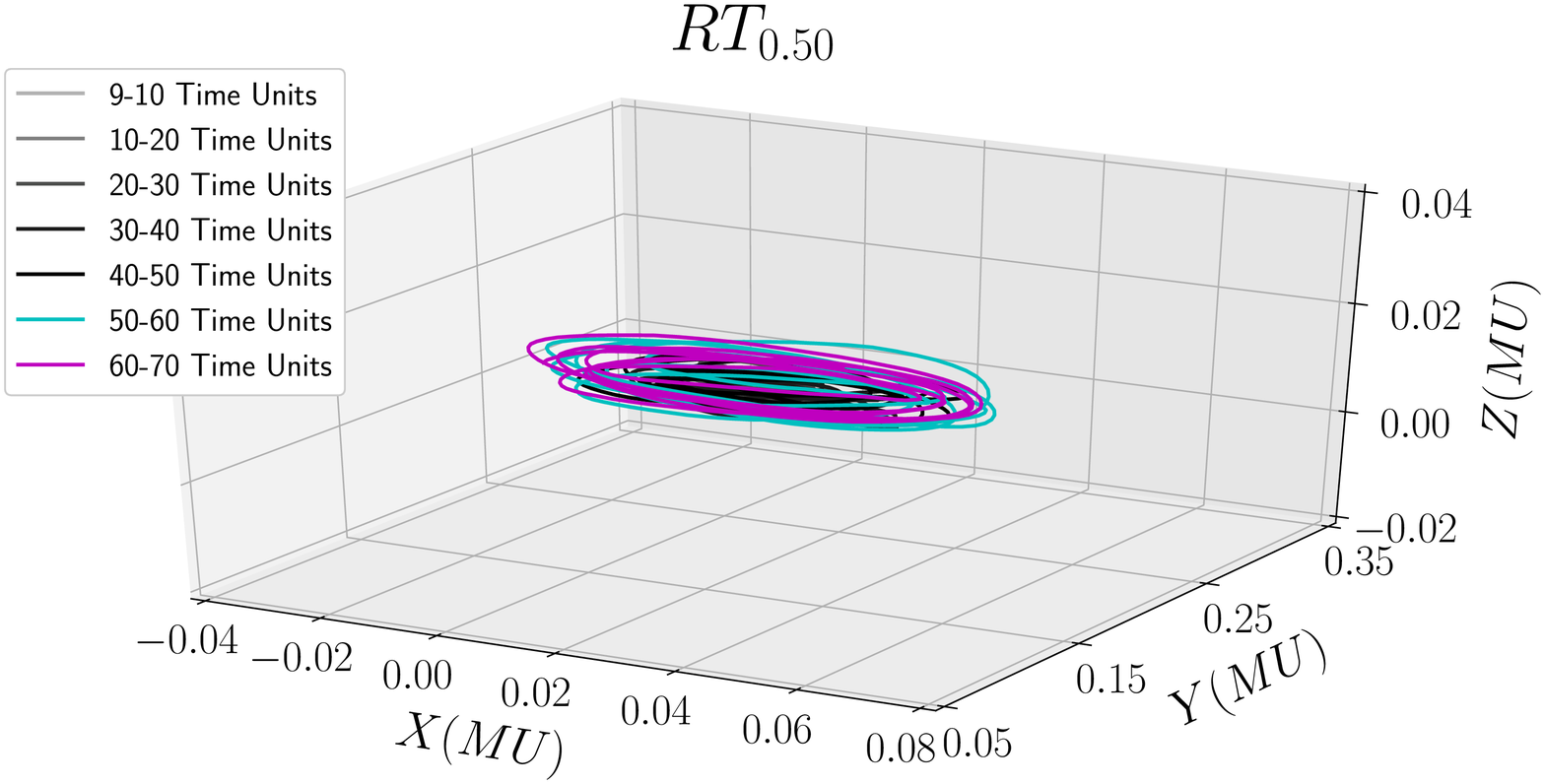}
		\includegraphics[width=1\linewidth]{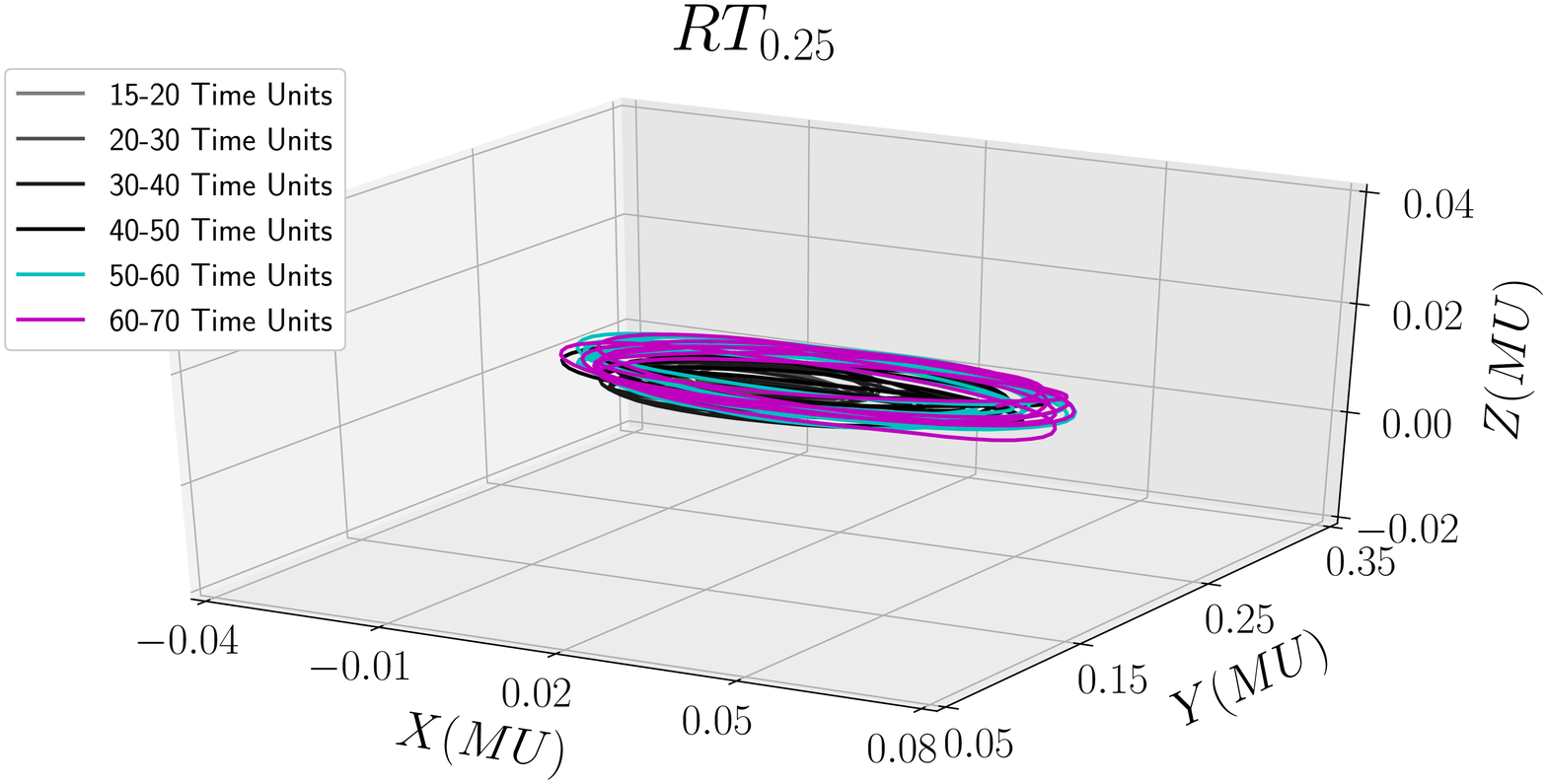}	
		\includegraphics[width=1\linewidth]{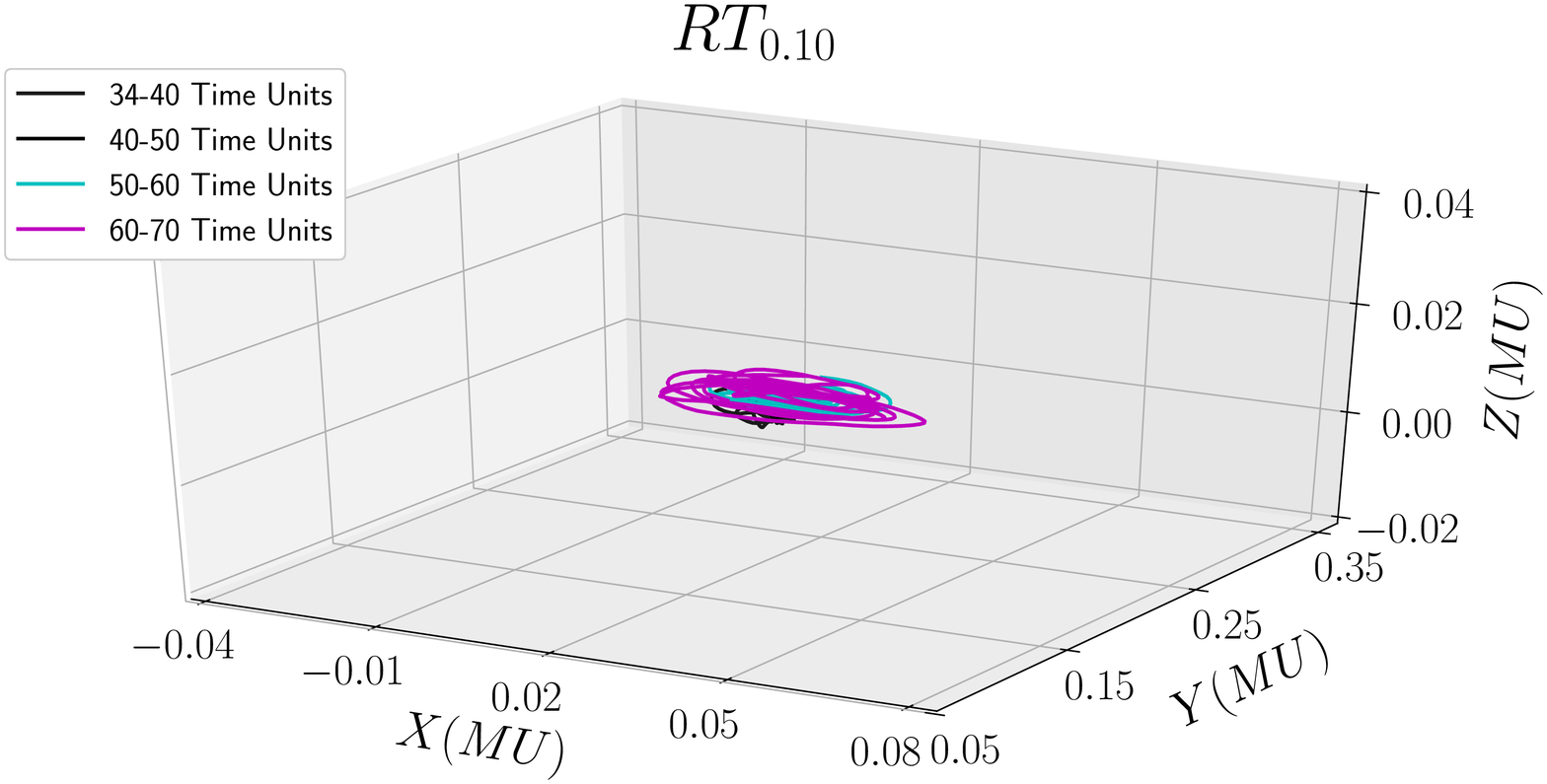}
		\includegraphics[width=1\linewidth]{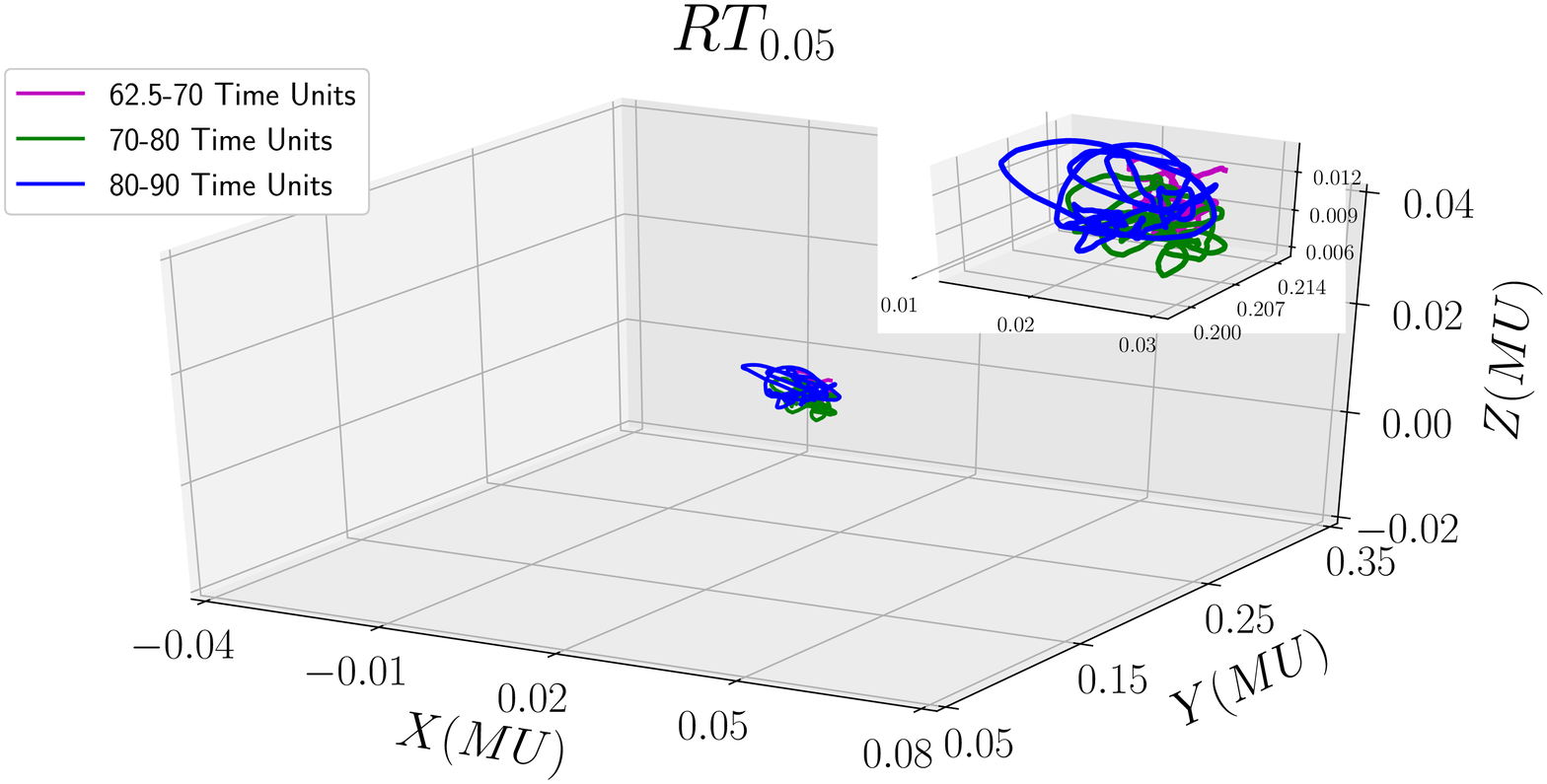}
		\caption{Motion of the SMBH binary center of mass for RT$_{1.00}$ (top) to RT$_{0.05}$ (bottom).The axes are fixed as discussed in the caption of Fig. \ref{fig:comRF}.}		
		\label{fig:comN}
	\end{figure}
	
	\begin{figure}
		\centering
		\includegraphics[width=1\linewidth]{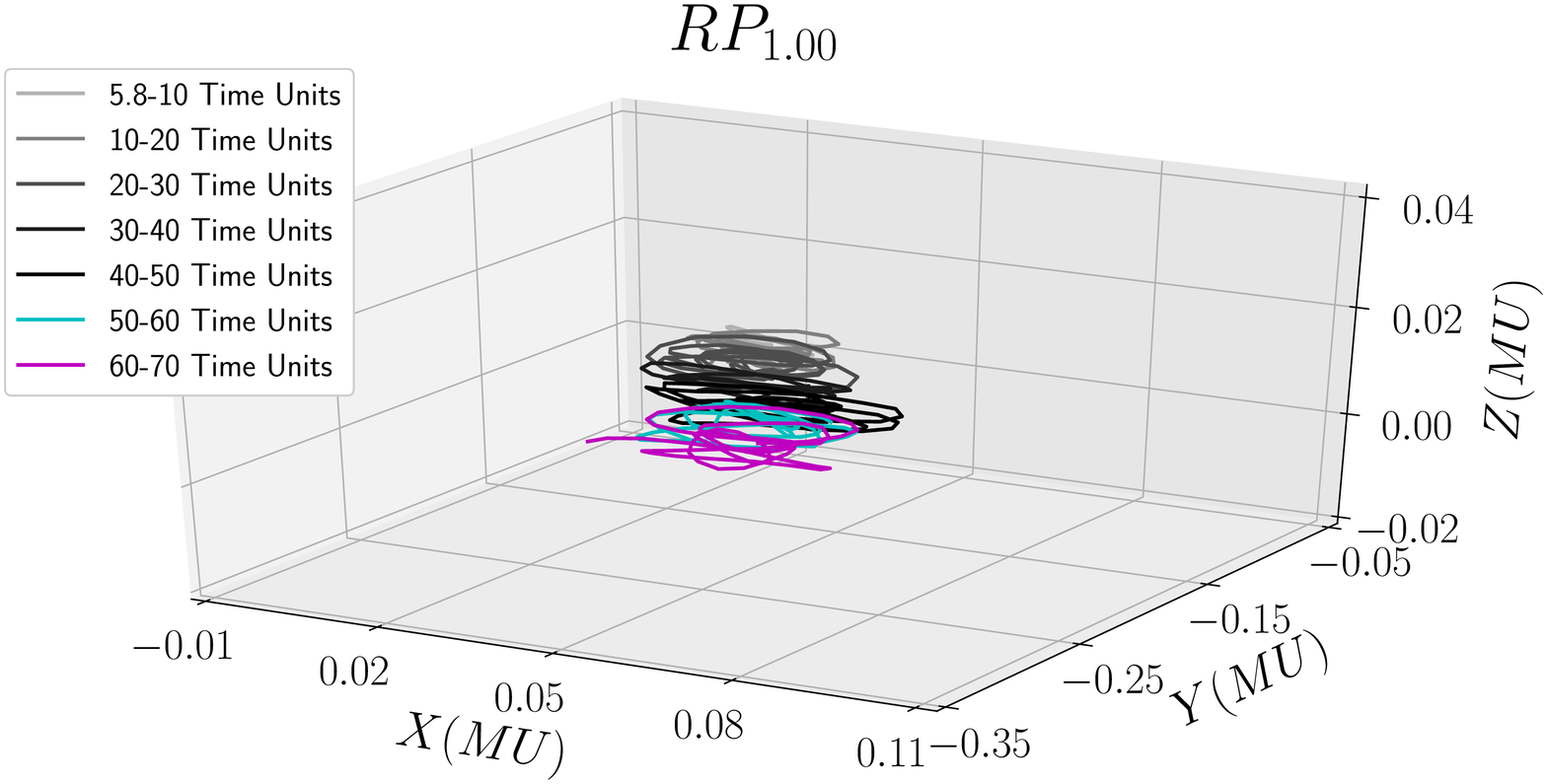}
		\includegraphics[width=1\linewidth]{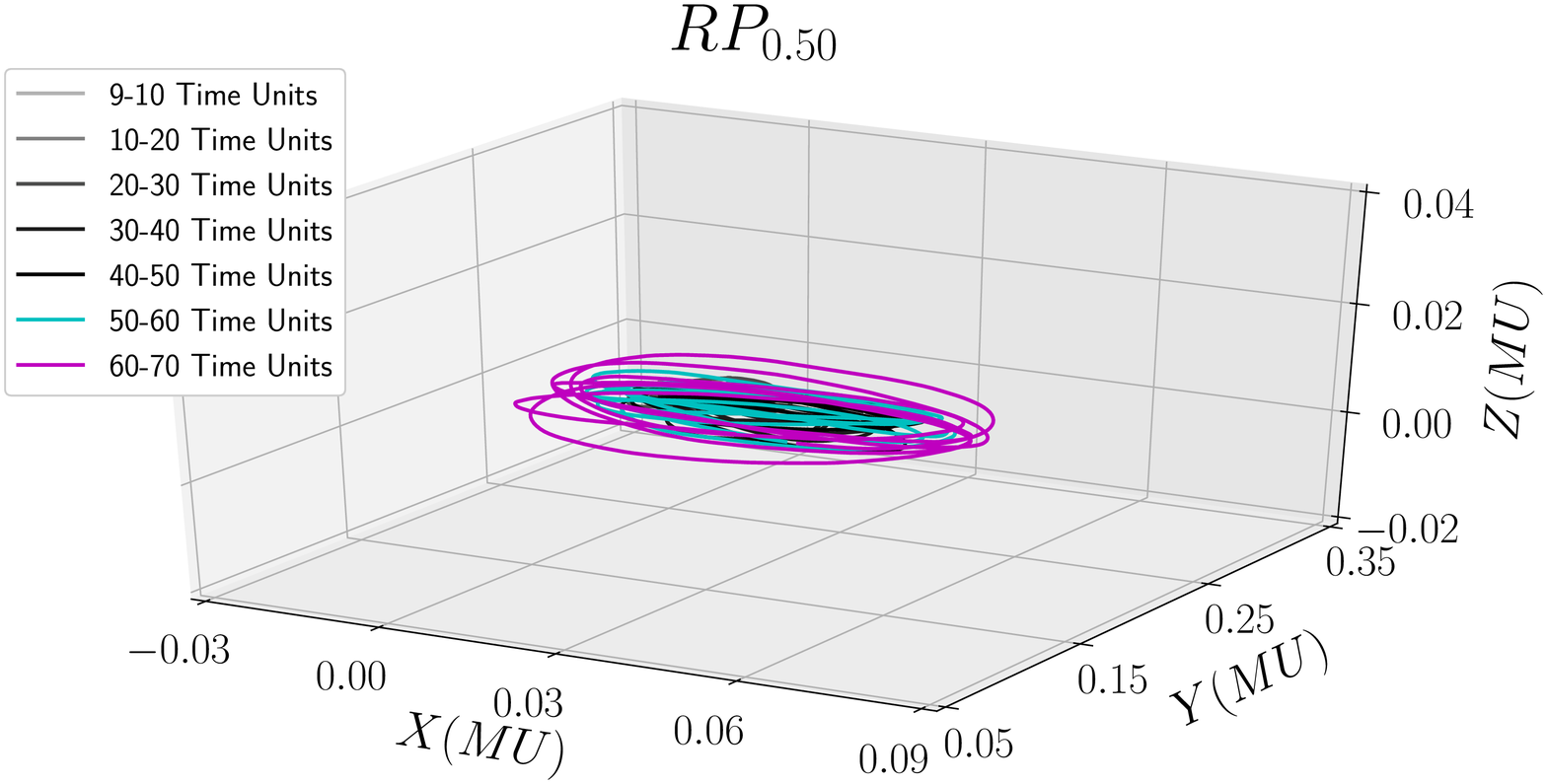}
		\includegraphics[width=1\linewidth]{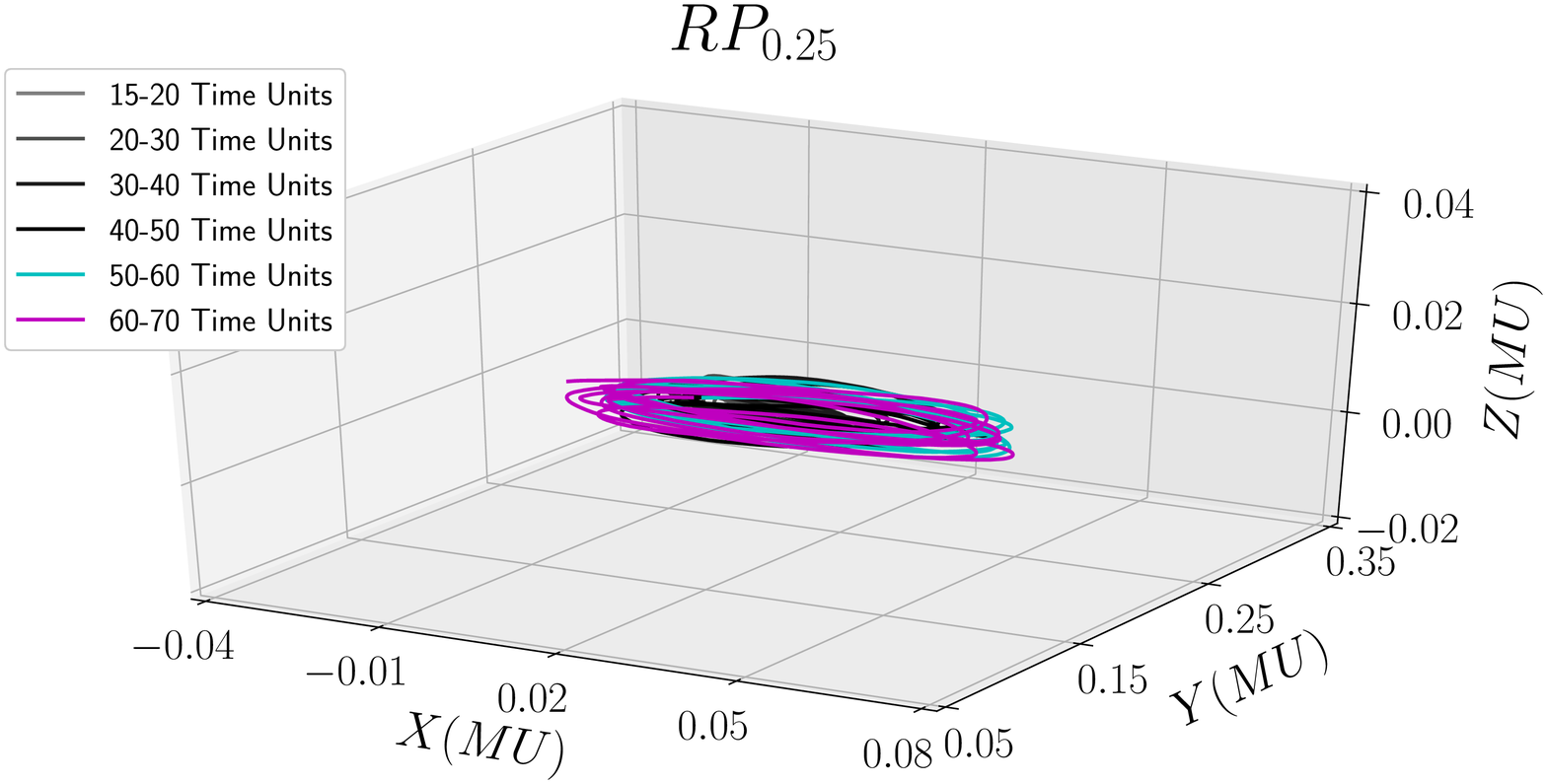}	
		\includegraphics[width=1\linewidth]{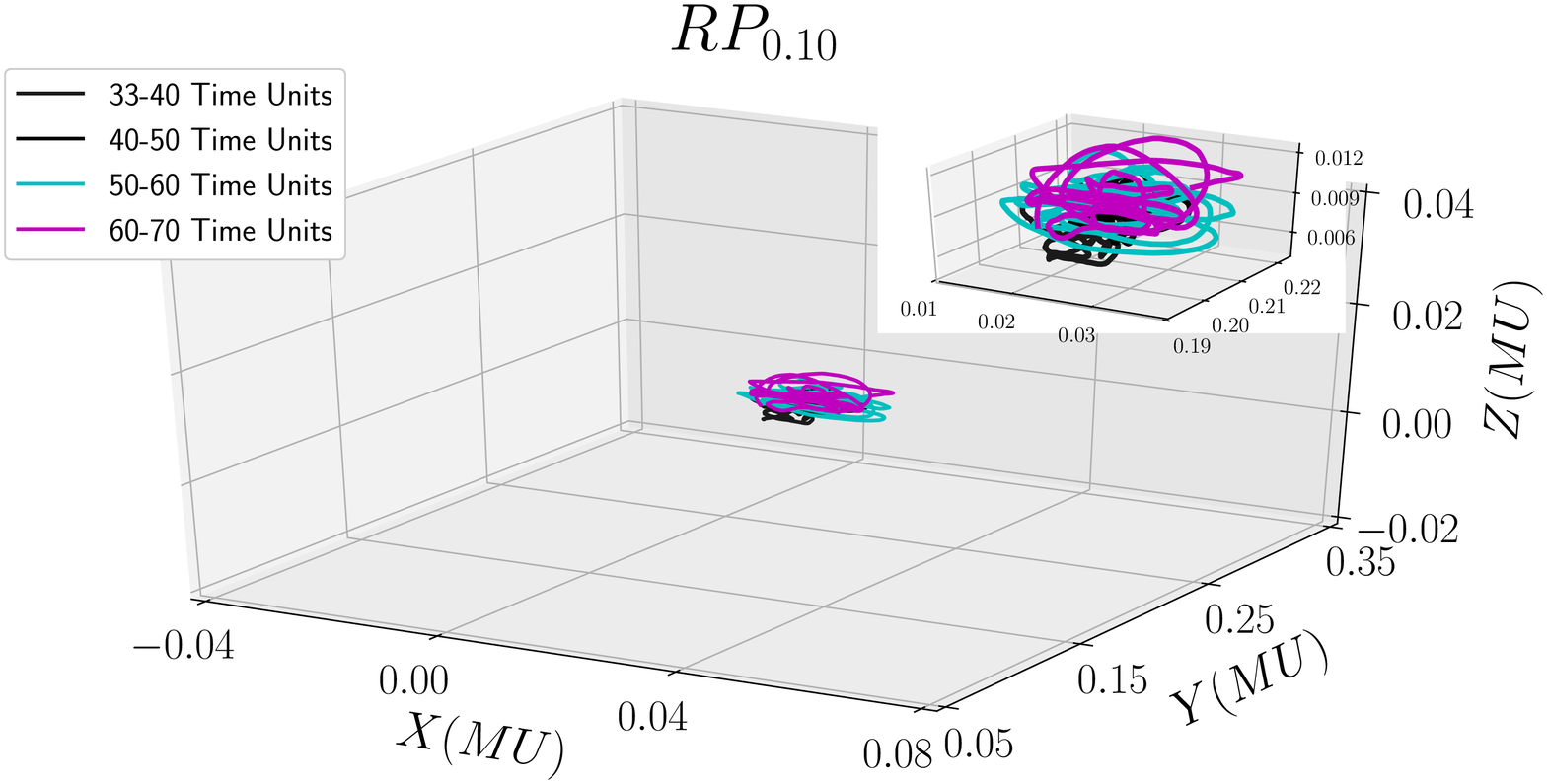}			
		\includegraphics[width=1\linewidth]{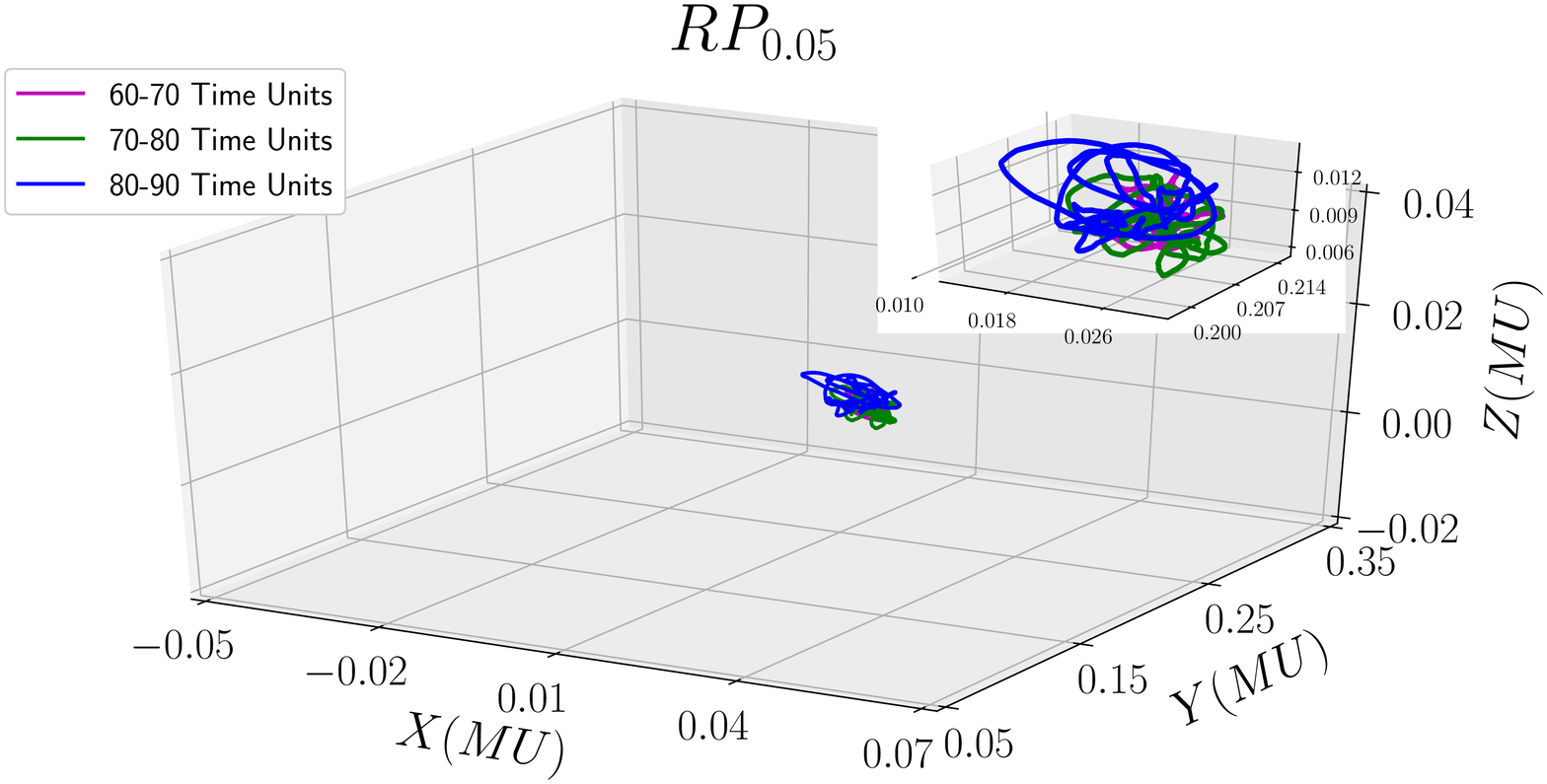}	 
		\caption{Motion of binary SMBH center of mass for \textbf{RP}$_{1.00}$ (top) to \textbf{RP}$_{0.05}$ (bottom). The scaling is fixed as discussed in the caption of Fig. \ref{fig:comRF}.}		
		\label{fig:comNP}
	\end{figure}

	Our next consideration is to investigate the SMBH binary orbit inside its host nuclei ( see Fig. \ref{fig:comRF} (N models), Fig. \ref{fig:comO} (\textbf{P} models), Fig. \ref{fig:comN} (\textbf{RT} models) and Fig. \ref{fig:comNP} (\textbf{RP} models)). For \textbf{P} models, the width of the COM orbit progressively diminishes moving from equal to small mass-ratio SMBHs. 
	
	The same trend has been observed for the \textbf{RT} and \textbf{RP} suites. However, the span of the orbit for these retrograde cases is smaller than in the prograde runs. Lacking galactic rotation, \textbf{N} binaries are seen to maintain their central position modulo Brownian motion.

\subsection{Gravitational Wave Emission}

\begin{figure}
		\centering
		\includegraphics[width=1\linewidth]{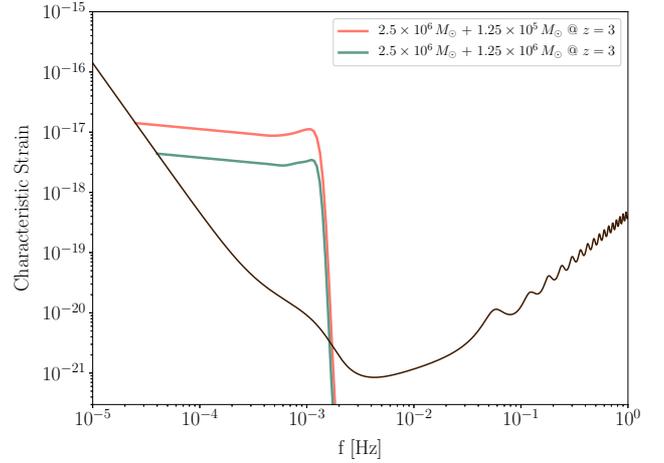}
		\caption{Gravitational wave signal from SMBH binary coalescence, as seen by LISA. Here we plot the $q=0.5$ and $q=0.05$ models within an M32 analog at redshift 3. Even with a very unequal mass ratio at high redshift, the merger is incredibly loud, with signal-to-noise ratios of ~850 and 300, respectively, for a 4-year LISA mission.}
		\label{fig:LISA}
	\end{figure}

Supermassive black hole binary coalescence produces the loudest gravitational wave signal in the known universe, making these systems a prime target for gravitational wave observatories. We calculated the expected gravitational wave signal from our models, assuming they were scaled to the galaxy M32 in mass and physical size. Taking this scaling yields the following unit conversions: 1 time unit = 0.11 Myr, 1 length unit = 0.03 kpc, and 1 mass unit = 5 $\times 10^8 M_{\odot}$. As a case study, we placed the SMBH merger itself at redshift 3, and determined both the strain and signal-to-noise ratio \citep{Neil} for the Laser Interferometer Space Antenna (LISA), a planned ESA/NASA gravitational wave observatory set to launch in 2034 (see Fig. \ref{fig:LISA}). It's useful to emphasize, though, that an SMBH merger at redshift 3 implies black hole pairing and galaxy merger times at, in some cases, much earlier epochs -- and as this study indicates, the timescale itself depends on the orientation of the binary orbital plane.  Assuming a constant hardening rate, and the time-averaged analytic expressions for semi-major axis and eccentricity evolution via gravitational wave emission from Peters and Matthews 
(1964), we estimated the binary evolution after the formal end of our $N$-body simulation. We note that this assumes the last part of the few-body scattering phase has no impact on the binary eccentricity, which is a safe assumption in the non-rotating case, but not true for rotation. Nonetheless, this paints a rough picture of the evolution unresolved by our runs.

	\section{Summary and Discussion} \label{summary}
	
	In the hierarchical galaxy formation framework, SMBH pairs are expected to form with a wide range of mass ratios ($q \sim 0.001 - 1$) in merger remnants~\citep{Dunn2018,Bellovary2019,KHB2010}. For mass ratios $q \leq 0.1$, the SMBH in the secondary galaxy is expected to be stripped of its surrounding stellar cusp and may wander at kiloparsec separations from the center of the primary galaxy \citep{cal11,Tremmel+18}, because the resulting dynamical friction timescales are thought to be long to form a bound system \citep{Dosopoulou+17}.
    
    In this study, we attempt to model the stellar dynamical evolution of relatively large mass ratio SMBH pairs with $q = 0.05 - 1$. We placed a secondary SMBH at a separation of about 10 times the black hole influence radius in both rotating and non-rotating flattened galactic nuclei models and mapped the orbital decay during the dynamical friction and few-body scattering phase. 
    
    \begin{itemize}
        
     \item  Dynamical friction depends on the kinematics of the stellar background. Prograde secondaries tend to decay faster and those in retrograde sink slower when versus those in non-rotating models. These discrepancies become even more evident for smaller mass ratios. This suggests that prograde infalling SMBHs pair and form a binary at lower mass ratios than those in non-rotating and retrograde orientations.
    
    \item    Orbits tend to circularise at binary formation time. They have low eccentricities in the rotation-free models. However, for $q < 0.1$, eccentricity begins to rise in hard binary regime, in line with \citet{Iwasawa+11}. Here we add that this mechanism of eccentricity growth seems to work best for smaller mass ratios ($q < 0.1$) as was also found in \citet{Khan+15}. For $q > 0.1$ SMBH binaries, this mechanism does not increase eccentricities efficiently.
    
    \item Prograde secondaries form binaries with low eccentricities ($e \sim 0.1$) and retain these low values in the hard binary regime through the end of our simulations. 
    
    \item Counter-rotation boosts eccentricities and binaries form with $e \sim 0.8$. For larger mass ratios, eccentricity grows further ($e \sim 0.9 - 0.99$), whereas for smaller mass ratios it drops down to intermediate values before exhibiting a slow and steady growth in the hard binary regime.
    
    \item In retrograde models, the secondary SMBH's orbital plane flips roughly around binary formation time, sometimes just before and sometime just after a binary forms according to our definition. We find that in most retrograde models in which the binary forms after the orbit flips, we end up with intermediate values of eccentricities due to moderate circularization in the loose binary phase. On the other hand, for cases in which the binary forms before the orbit flip, we reach and retain high eccentricities, as is expected in previous studies \citep{Sesana+11,holley+15,Mirza+17}. Subsequent forced flips do not cause a notable decrease in eccentricity as the stellar cusp surrounding the binary is already eroded.
    
    \item In a perfectly rotating galactic nuclei, all retrograde binaries flip their angular momentum to prograde before they reach a hard binary regime. However, how the degree of rotation impacts this conclusion for different mass ratios needs further investigation, which we are currently pursuing. 
    
    \item The SMBH binary hardening rates $s$ in the few-body scattering regime are very similar for both co and initially counter-rotating models, as the latter also flips to co-rotation around the few-body scattering phase. $s$ in these models are roughly (20-30) percent higher than their non-rotating counterparts. These higher hardening rates may be attributed to large span of the binary's center of mass,  enabling the binary to draw interactions from a larger phase space reservoir.
    
    \item In rotating galaxy models, the center of mass of the SMBH binary begins a small co-rotating orbit which, incredibly, expands swiftly to a large size while the separation between the black holes themselves shrink to a hard binary phase. We suspect that stellar scattering during the loose binary phase erodes the stellar density cusp and hence the underlying potential, which in turn causes SMBH binary center of mass to drift out. The center of mass orbit then stabilizes during the hard binary phase, with a spatial extent for $q = 0.25 - 1.0$ that is of the order of the binary's influence radius. We predict that SMBH binaries may be offset by a few parsecs to a few tens of parsecs from the centers of their host galaxies, and in motion, in highly rotating nuclei.
    
    \end{itemize}
    
    In this study, we investigated aspects of SMBH binary evolution in rotating host galactic nuclei. However, this study considered a perfectly flattened galaxy with a single bulk rotation speed and a single density profile. In a series of follow-up studies, we are exploring SMBH binary dynamics in a broad sample of ever more realistic background models. Our ultimate goal is to provide a formalism to estimate the SMBH decay time and eccentricity as a function of galaxy structure, kinematics, morphology, and SMBH orbital parameters. 
    \section*{Acknowledgments}
    
    We acknowledge the support by Vanderbilt University for providing access to its Advanced Computing Center for Research and Education (ACCRE). 
    FK and KHB were supported through NASA ATP Grant 80NSSC18K0523.
    FK acknowledges the support of Higher Education Commission of Pakistan through National Research Program for Universities (NRPU) project 4159.
    MAM would like to thank Bonn-Cologne Graduate School of Physics and Astronomy, Germany for providing funding in the framework of the honors branch.

	%%%%%%%%%%%%%%%%%%%%%%%%%%%%%%%%%%%%%%%%%%%%%%%%%%
	
	%%%%%%%%%%%%%%%%%%%% REFERENCES %%%%%%%%%%%%%%%%%%

	\bibliographystyle{mnras}
	\bibliography{ms}

	% Don't change these lines
	\bsp	% typesetting comment
	\label{lastpage}
\end{document}